\documentclass[12pt]{article}

\usepackage{amsmath}
\usepackage{amssymb}
\usepackage{siunitx}

\usepackage{booktabs}
\usepackage{graphicx}
\usepackage{caption}
\usepackage{multirow}
\usepackage{algorithm}
\usepackage{algpseudocode}

\usepackage{geometry}
\usepackage{textcomp}

\usepackage[numbers]{natbib}

\usepackage[colorlinks=true,
            linkcolor=black,
            citecolor=black,
            urlcolor=blue]{hyperref}
\usepackage{orcidlink}

\usepackage{setspace}
\usepackage{authblk}

\newcommand{\weik}{w_{\mathrm{eik}}}

\newcommand{\phih}{\hat{\phi}}
\newcommand{\Lpde}{\mathcal{L}_{\mathrm{pde}}}
\newcommand{\Lic}{\mathcal{L}_{\mathrm{ic}}}
\newcommand{\Leik}{\mathcal{L}_{\mathrm{eik}}}

\begin{document}

\title{SDF-Aware Weighting: Adaptive Eikonal Regularisation for
       Three-Dimensional Level-Set\\
       Physics-Informed Neural Networks}

\author[1]{Muhammad Akbar Khan~\orcidlink{0009-0001-7956-0080}%
\thanks{Email: \href{mailto:akbar.bsma1337@gmail.com}{akbar.bsma1337@gmail.com}}}
\affil[1]{Department of Mathematics, NED University of Engineering and
          Technology, Karachi 75270, Pakistan}

\date{}

\maketitle

\begin{abstract}
\noindent
Adaptive loss-balancing schemes for physics-informed neural networks rest on a
premise that every residual in the composite loss should be driven to zero, so
that the weights need only equalise the terms' contributions to the update. For
level-set advection with an eikonal regulariser that premise fails. The eikonal
term penalises the deviation of $\lVert\nabla\phi\rVert$ from unity, a property
that transport preserves only under rigid motion of a smooth interface; where
the exact solution departs from a signed-distance function the eikonal residual
of the correct answer is nonzero, and a scheme that drives it towards zero moves
the network away from that answer. We show that standard gradient-norm balancing
fails in exactly this way, its weight remaining at or near its initial value for
the whole of training on the benchmarks where the property is violated, and we
introduce SDF-Aware Weighting (SAW), which
combines a residual-quantile gate with a gradient-norm ratio so that points
exhibiting legitimate departure are excluded before the surviving term is
scaled. Across four three-dimensional benchmarks SAW selects an eikonal weight
within an order of magnitude of the value located by an eighteen-run manual sweep,
spanning four decades from $10^{-1}$ to $10^{-5}$ with a single fixed
configuration. On the slotted sphere, where the constructive-solid-geometry
initial field is non-differentiable at the reentrant edges of the slot, SAW
attains a lower error than any weight in that sweep. Two smooth
rigid benchmarks serve as controls: their exact solutions remain
signed-distance functions for all time, the gate has no legitimate violations to
suppress, and SAW is correspondingly worse than a fixed weight there: the
mechanism does not fire when its premise does not hold. An ablation with the
gate disabled isolates its contribution: without it the slot is all but
entirely filled by the end of the integration while the relative $L_2$ error
reads
$1.06\%$, indistinguishable from a field that never represented the slot at all.
We therefore give a feature-restricted measure that separates the two cases.
\end{abstract}

\vspace{2mm}

\vspace{4mm}

\section{Introduction}
\label{sec:intro}

The level-set method represents a moving interface implicitly, as the zero
contour of a higher-dimensional function $\phi$ transported by the flow
\citep{Osher1988}. Because the interface is never meshed, topological changes
such as merging and breakup require no special treatment, which is why the
formulation is used throughout multiphase flow, combustion, solidification and
image segmentation. In the simplest case the interface is
carried by a prescribed velocity field and $\phi$ satisfies a linear transport
equation.

That equation is deceptively difficult to solve accurately. The quantity of
interest is not $\phi$ itself but its zero level set, and a scheme may reduce
the error in $\phi$ while still losing volume, rounding corners, or destroying
thin features. It is standard practice to initialise $\phi$ as a signed
distance function (SDF), for which $\lVert\nabla\phi\rVert = 1$, because the
conditioning of the field near the interface then does not degrade; but this
property is not preserved by transport except under rigid motion, and classical
solvers restore it by periodically re-solving a separate reinitialisation
equation \citep{Sussman1994, SussmanFatemi1999}.

Physics-informed neural networks~\citep{Raissi2019} (PINNs) offer an alternative in
which $\phi$ is represented by a network trained to satisfy the governing
equation and its initial data in a least-squares sense at sampled collocation
points. The representation is mesh-free and continuous in space and time, so
there is no Courant--Friedrichs--Lewy (CFL) restriction and no interpolation is required to evaluate the
solution between grid points or time levels. Against this, training is a
non-convex optimisation whose outcome depends on architecture, sampling and, in
particular, on the relative weighting of the terms in the composite loss.
\citet{Krishnapriyan2021} showed that physics-informed networks can fail on
convection-dominated problems even when the network is expressive enough to
represent the solution, so the difficulty lies in the optimisation rather than
in the approximation. Level-set advection is convection dominated by
construction, which is why the configuration of the training problem, and the
loss weighting in particular, matters as much as it does here.

\subsection{Physics-informed neural networks for level-set advection}

Several studies have applied physics-informed networks to interface
transport. \citet{Mullins2025} examined several architectures for moving
interface problems in the level-set formulation and observed that adding an
eikonal regularisation term to the loss improves the results when given an
appropriate weight; they did not characterise what that weight should be.
\citet{Bi2025} developed an extended-interface formulation for
moving interface problems. More recently, \citet{Zhai2026} proposed a meshfree
piecewise-network formulation for two-phase flows with prescribed or
solution-driven interface motion, with rigorous approximation and error
estimates, though without an explicit level-set representation, and
\citet{Chang2025} coupled a dedicated interface-tracking network with a
modified zero-level-set function to resolve the Stefan problem, including
Mullins--Sekerka instabilities. In previous work we conducted a systematic ablation
of network and loss configurations for two-dimensional level-set advection
across four benchmarks \citep{Khan2026}, establishing an architecture and
training protocol and identifying the eikonal weight as the dominant
hyperparameter. That study located the weight by sweep, and closed by naming the
cost of doing so as the principal obstacle to applying the method without manual
intervention.

Two threads of the wider physics-informed literature bear directly on the
present work. The first concerns \emph{where} the residual is enforced:
residual-based adaptive distribution and refinement concentrate collocation
points where the equation is least well satisfied \citep{Wu2023, Lu2021}, which
matters when the solution has localised features occupying a small fraction of
the domain. The second concerns \emph{how} the loss terms are balanced. Manually
chosen weights are a known source of failure \citep{Wang2021grad,
Krishnapriyan2021}, and several schemes adapt them during training, whether by
gradient statistics \citep{Maddu2022}, by trainable pointwise multipliers
\citep{McClenny2023}, or by uncertainty \citep{Xiang2022}. Adaptive balancing,
however, presupposes that the terms are mutually compatible, so that driving all
of them down is the correct objective. The eikonal term does not always satisfy
that premise. Where the exact solution departs from a signed-distance function
(near the reentrant edges of a constructive-solid-geometry initial field, or
throughout a stretched interface) the eikonal residual of the correct answer
is nonzero. A scheme that reads a large residual as network error and increases
the pressure accordingly is then moving in the wrong direction, and none of the
rules above distinguishes the two cases. This paper identifies that failure,
shows that it is structural rather than a matter of tuning, and repairs it with a
residual-quantile gate that removes the offending points before the weight is
set.

\subsection{Contributions}

\begin{enumerate}
    \item \textbf{Adaptive loss balancing fails when a residual should not
    vanish.} We show that gradient-norm balancing applied to the eikonal term
    leaves the weight at or near its initial value for the whole of training on
    the benchmarks where the signed-distance property fails, because the term it
    is balancing stays small
    precisely where the exact solution departs from a signed-distance function.
    The failure is not a tuning problem: it is a consequence of the premise the
    method rests on.

    \item \textbf{SDF-Aware Weighting (SAW).} A residual-quantile gate excludes
    points whose eikonal residual is evidence of legitimate departure rather
    than of network error, after which a gradient-norm ratio scales the
    surviving term. The gate is what allows the ratio to move at all.

    \item \textbf{The selected weight tracks the problem across four decades.}
    With one fixed configuration SAW selects a weight within an order of
    magnitude of the value located by an eighteen-run manual sweep on every
    benchmark,
    from $10^{-1}$ on a rigidly transported smooth sphere to $10^{-5}$ under
    deformation.

    \item \textbf{On the benchmark it targets, SAW improves on the sweep it
    replaces.} For the slotted sphere it attains a lower error than any of the
    six weights the sweep examined, in a single run rather than eighteen.

    \item \textbf{Negative controls.} On two benchmarks whose exact solutions
    remain signed-distance functions for all time, the gate removes correct
    supervision and SAW is worse than a fixed weight. We report this: a method
    that improved every case would be evidence of a generic regularisation
    effect rather than of the mechanism claimed.

    \item \textbf{Global norms cannot certify thin features.} On the slotted
    sphere, replacing the exact solution by a sphere with no slot at all changes
    the relative $L_2$ error by about one percentage point, so that measure
    cannot distinguish a captured slot from an absent one. We give a sign-based,
    feature-restricted measure that can.
\end{enumerate}

Section~\ref{sec:method} states the formulation, the network and the training
procedure. Section~\ref{sec:benchmarks} defines the four benchmarks and the
error measures. Section~\ref{sec:eik} reports the fixed-weight sweeps that establish the target,
Section~\ref{sec:saw:results} the SAW results and the gate ablation.
Section~\ref{sec:baseline} gives the comparison against the
classical solver, Section~\ref{sec:discussion} discusses limitations, and
Section~\ref{sec:conclusion} concludes.

\section{Methodology}
\label{sec:method}

\subsection{Level-set formulation and loss function}

The interface is represented implicitly as the zero contour of a level-set
field transported by a prescribed velocity,
\begin{equation}
    \frac{\partial \phi}{\partial t} + \mathbf{u}\cdot\nabla\phi = 0 ,
    \qquad \mathbf{x}\in\Omega = [0,1]^3,\; t\in[0,T],
    \label{eq:transport}
\end{equation}
with $\phi(\mathbf{x}, 0) = \phi_0(\mathbf{x})$ the signed distance function of
the initial interface.

\begin{figure}[htbp]
\centering
\includegraphics[width=\textwidth]{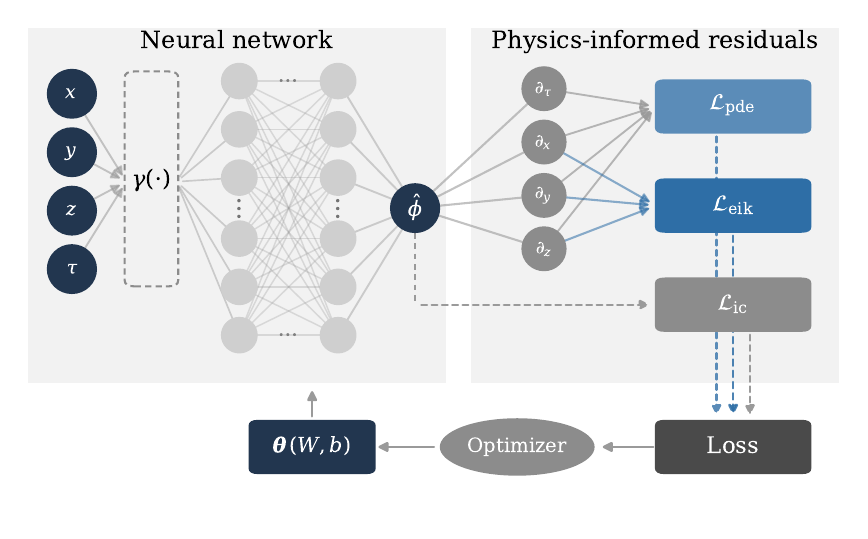}
\caption{Network architecture and training loop. The dashed encoding block
$\gamma(\cdot)$ is used only for the slotted-sphere benchmark. Loss terms are
defined in Equation~\eqref{eq:loss}; the residual is
Equation~\eqref{eq:residual}.}
\label{fig:architecture}
\end{figure}

The level-set field is approximated by a neural network
$\hat{\phi}(\mathbf{x}, \tau; \theta)$ with parameters $\theta$, taking as input
the three spatial coordinates and a \emph{normalised} time
$\tau = t/T \in [0,1]$. Normalisation keeps all four inputs on a comparable
scale, and introduces an explicit factor of $T$ in the residual through the
chain rule: writing \eqref{eq:transport} in terms of $\tau$ gives
\begin{equation}
    r(\mathbf{x}, \tau; \theta)
    = \frac{\partial\hat{\phi}}{\partial\tau}
      + T\,\mathbf{u}(\mathbf{x}, \tau)\cdot\nabla\hat{\phi} .
    \label{eq:residual}
\end{equation}
All derivatives are obtained by automatic differentiation, so no spatial or
temporal discretisation enters the solution procedure.

Training minimises the composite objective
\begin{equation}
    \mathcal{L}
    = w_{\mathrm{pde}}\,\Lpde
    + w_{\mathrm{ic}}\,\Lic
    + \weik\,\Leik ,
    \label{eq:loss}
\end{equation}
whose three terms are the mean squared residual, the mismatch against the
prescribed initial field, and the eikonal regulariser:
\begin{align}
    \mathcal{L}_{\mathrm{pde}}
    &= \frac{1}{N_f}\sum_{i=1}^{N_f} r(\mathbf{x}_i, \tau_i; \theta)^2 ,\\
    \mathcal{L}_{\mathrm{ic}}
    &= \frac{1}{N_i}\sum_{j=1}^{N_i}
       \bigl(\hat{\phi}(\mathbf{x}_j, 0; \theta) - \phi_0(\mathbf{x}_j)\bigr)^2 ,\\
    \mathcal{L}_{\mathrm{eik}}
    &= \frac{1}{N_f}\sum_{i=1}^{N_f}
       \bigl(\lVert\nabla\hat{\phi}(\mathbf{x}_i, \tau_i; \theta)\rVert - 1\bigr)^2 .
\end{align}
No boundary condition is imposed: in every benchmark the interface remains
interior to the domain for all time, and the network is free to determine the
far field. Throughout, $w_{\mathrm{pde}} = 1$ and $w_{\mathrm{ic}} = 10$; the
determination of $\weik$ is the subject of Section~\ref{sec:eik}. Component
losses are recorded unweighted, so that runs at different $\weik$ remain
comparable term by term. Figure~\ref{fig:architecture} summarises the
architecture and the training loop.

\subsection{Neural-network architecture}

The network is a fully connected feed-forward architecture of eight hidden
layers with $256$ units each and $\tanh$ activations, initialised by the Xavier
scheme~\citep{Glorot2010}, following the configuration established for the
two-dimensional
benchmarks~\citep{Khan2026}.

For the slotted sphere the raw coordinates are first mapped through a random
Fourier feature encoding~\citep{Tancik2020},
\begin{equation}
    \gamma(\mathbf{v}) = \bigl[\sin(\mathbf{v}^{\top}B),\;
                               \cos(\mathbf{v}^{\top}B)\bigr],
    \qquad B_{k\ell} \sim \mathcal{N}(0, \sigma^2),
\end{equation}
with $128$ features and $\sigma = 5$, which mitigates the spectral bias of
fully connected networks towards low frequencies. The three smooth-interface
benchmarks do not use the encoding, since their solutions contain no
comparably sharp features.

\subsection{Collocation and adaptive sampling}

Collocation points are drawn uniformly in the four-dimensional domain
$\Omega\times[0,1]$ and initial-condition points uniformly in $\Omega$ at
$\tau = 0$. For the slotted sphere and the reversed vortex (defined in Section~\ref{sec:benchmarks}) the collocation set
is additionally adapted during training by two mechanisms. Residual-based
adaptive distribution~\citep{Wu2023} (RAD) resamples the base set every $10^{3}$
iterations from a pool of $2\times10^{4}$ candidates, with selection
probability proportional to $|r|^{k}/\overline{|r|^{k}} + c$ using $k = c = 1$.
Residual-based adaptive refinement~\citep{Lu2021} (RAR) additionally appends the $500$
highest-residual points from a pool of $5\times10^{4}$ every $1800$ iterations,
so the collocation set grows over training. The initial-condition points are
not adapted.

\subsection{Causal weighting}

For the two benchmarks that use adaptive sampling, the residual term is
weighted causally~\citep{Wang2022}. The collocation points are sorted by $\tau$
and partitioned into $M$ contiguous chunks; chunk $m$ receives the weight
\begin{equation}
    \omega_m = \exp\Bigl(-\epsilon \sum_{k<m}
                \overline{r^2}_k\Bigr),
\end{equation}
where $\overline{r^2}_k$ is the mean squared residual of chunk $k$, held fixed
with respect to the gradient. Later times are therefore down-weighted until
earlier ones are well satisfied, which discourages the network from fitting a
solution that is accurate late in the interval but inconsistent with the
initial data. We use $\epsilon = 1$ throughout, with $M = 32$ for the slotted
sphere and $M = 10$ for the reversed vortex.

\subsection{SDF-Aware Weighting}
\label{sec:saw}

The eikonal weight $\weik$ is the one quantity in \eqref{eq:loss} that the
problem statement does not fix, and its optimum spans four decades across the
benchmarks considered here. SDF-Aware Weighting derives it from the training
dynamics instead. The scheme has two parts, applied in order at every
iteration.

\paragraph{Residual-quantile gate.}
Let $r_i = \bigl(\lVert\nabla\phih(\mathbf{x}_i,\tau_i)\rVert - 1\bigr)^2$ be
the pointwise eikonal residual. Adaptive balancing normally treats a large
$r_i$ as evidence that the network is failing there and increases the pressure
accordingly. For level-set advection that inference is unsound: where the exact
solution is not a signed-distance function, $r_i$ is large because the correct
answer makes it large. We therefore exclude the upper tail. Writing $\tau_q$
for the $q$-th quantile of $\{r_i\}$, smoothed across iterations by
\begin{equation}
    \hat{\tau} \leftarrow 0.99\,\hat{\tau} + 0.01\,\tau_q ,
\end{equation}
each point receives the binary weight $m_i = \mathbf{1}[r_i \le \hat{\tau}]$,
and the eikonal loss becomes the mean of $m_i r_i$. The gate is deliberately
hard rather than a soft down-weighting: a point either exhibits a legitimate
departure from the signed-distance property or it does not, and a graded
response would reintroduce pressure at exactly the points the gate exists to
protect.

\paragraph{Gradient-norm ratio.}
The surviving term is then scaled so that its influence on the parameter update
matches that of the other two. With
$g_k = \lVert\nabla_{\!\theta}\mathcal{L}_k\rVert$ evaluated on the final
layer's parameters, we form
\begin{equation}
    \weik^{\mathrm{raw}}
    = \mathrm{clip}\!\left(\frac{g_{\mathrm{pde}} + g_{\mathrm{ic}}}
                                {g_{\mathrm{eik}} + \varepsilon},\; 0,\;
                           w_{\mathrm{pde}}\right),
    \qquad
    \weik \leftarrow \beta\,\weik + (1-\beta)\,\weik^{\mathrm{raw}} ,
    \label{eq:saw_ratio}
\end{equation}
with $\varepsilon = 10^{-8}$, distinct from the causal-weighting parameter
$\epsilon$ of Section~\ref{sec:method}, and $\beta = 0.999$. The clip is applied before the
exponential moving average rather than after: as training converges
$g_{\mathrm{eik}}$ approaches zero and the raw ratio grows without bound, and an
unclamped value entering the average would take $\mathcal{O}(1/(1-\beta))$
iterations to unwind. Restricting the norms to the final layer is the standard
proxy in this literature, and it reduces the cost of the three additional
backward passes that the ratio requires.

\paragraph{The two parts are not independent.}
It is the gate that allows the ratio to move. With the gate disabled the
eikonal loss remains small throughout, $g_{\mathrm{eik}}$ stays correspondingly
small, the ratio stays at or near its clip, and $\weik$ fails to descend to the
value a sweep identifies, remaining at $w_{\mathrm{pde}}$ on two benchmarks
and reaching only $0.70$ on a third. On the two benchmarks where the
signed-distance property fails these are close to the worst settings available.
Section~\ref{sec:saw:ablation} reports this ablation. SAW therefore replaces one hyperparameter whose optimum
varies by four decades with two, $\beta$ and $q$, whose settings we hold fixed
across every benchmark. Algorithm~\ref{alg:saw} states the scheme; the
surrounding training loop is unchanged, so SAW attaches to any protocol that
supplies the three loss terms.

\begin{algorithm}[htbp]
\caption{SDF-Aware Weighting: the eikonal weight for one iteration}
\label{alg:saw}
\begin{algorithmic}[1]
\Require pointwise eikonal residuals $\{r^{\mathrm{eik}}_i\}_{i=1}^{N_f}$;
         losses $\Lpde$, $\Lic$; weights $w_{\mathrm{pde}}$, $w_{\mathrm{ic}}$;
         quantile $q$; smoothing $\beta$; state $\hat{\tau}$, $\weik$ from the
         previous iteration
\Ensure  the eikonal contribution to $\mathcal{L}$
\Statex
\Statex \textbf{Gate: exclude the points whose residual the exact solution makes large}
\State $\hat{\tau} \gets 0.99\,\hat{\tau}
        + 0.01\,\mathrm{quantile}_q\bigl(\{r^{\mathrm{eik}}_i\}\bigr)$
\State $m_i \gets \mathbf{1}\bigl[r^{\mathrm{eik}}_i \le \hat{\tau}\bigr]$
\State $\Leik \gets \overline{m_i\,r^{\mathrm{eik}}_i}$
\Statex
\Statex \textbf{Ratio: scale what survives to match the other terms}
\State $g_k \gets \lVert\nabla_{\!\theta}\mathcal{L}_k\rVert$ on the final
        layer, for $k \in \{\mathrm{pde}, \mathrm{ic}, \mathrm{eik}\}$
\State $\weik^{\mathrm{raw}} \gets \mathrm{clip}\bigl(
        (g_{\mathrm{pde}} + g_{\mathrm{ic}})/(g_{\mathrm{eik}} + \varepsilon),
        \; 0,\; w_{\mathrm{pde}}\bigr)$
\State $\weik \gets \beta\,\weik + (1-\beta)\,\weik^{\mathrm{raw}}$
\Statex
\State \Return $\weik\,\Leik$
\end{algorithmic}
\end{algorithm}

\subsection{Optimisation and training procedure}

Training proceeds in two stages. Adam~\citep{Kingma2015} is run for
$2\times10^{4}$ iterations at
an initial learning rate of $10^{-3}$ under a cosine annealing
schedule~\citep{Loshchilov2017} completing one full cycle to $\eta_{\min} = 10^{-5}$, with gradient norms
clipped at unity. Limited-memory Broyden--Fletcher--Goldfarb--Shanno (L-BFGS) with a strong-Wolfe line search then refines the
result. The second stage is retained as an optional refinement rather than a
required one: where the annealed Adam iterate is already tightly converged the
line search finds no admissible step and the stage is inactive, whereas where
convergence is incomplete it reduces the loss substantially. Both outcomes
occur across the benchmarks reported here, and the improvement attributable to
L-BFGS is recorded for every run.

Random seeds control both the network initialisation and the collocation
sampling, so the seed-to-seed spread reported in Section~\ref{sec:eik} covers
both sources of variability. Table~\ref{tab:config} collects the settings that
differ between benchmarks.

\begin{table}[htbp]
\centering
\caption{Configuration by benchmark (labels defined in Section~\ref{sec:benchmarks}, which follows). 
Settings not listed are common to all
four: eight hidden layers of $256$ $\tanh$ units, $w_{\mathrm{pde}} = 1$,
$w_{\mathrm{ic}} = 10$, Adam for $2\times10^{4}$ iterations under a cosine
schedule from $10^{-3}$ to $10^{-5}$, and evaluation on a $201^3$ grid at six
snapshots.}
\label{tab:config}
\begin{tabular}{lcccc}
\toprule
 & TR3D & RO3D & ZD3D & RV3D \\
\midrule
$T$                       & $5$   & $2\pi$ & $2\pi$ & $2$ \\
$N_f$                     & $10^4$ & $10^4$ & $2\times10^4$ & $10^4$ \\
$N_i$                     & $5\times10^3$ & $5\times10^3$ & $10^4$ & $5\times10^3$ \\
Fourier features          & --    & --     & $128$, $\sigma=5$ & -- \\
RAD / RAR                 & --    & --     & yes & yes \\
causal chunks $M$         & --    & --     & $32$ & $10$ \\
L-BFGS iterations         & $500$ & $500$  & $500$ & $2000$ \\
selected $\weik$          & $10^{-1}$ & $10^{-1}$ & $10^{-3}$ & $10^{-5}$ \\
\bottomrule
\end{tabular}

\vspace{2mm}
\begin{minipage}{0.92\textwidth}
\footnotesize The ZD3D sweep was run at both sampling budgets and the same
weight ($10^{-3}$) is selected at each; the reported configuration uses
$N_f = 2\times10^{4}$, $N_i = 10^{4}$. See Table~\ref{tab:fixed_sweeps}.
\end{minipage}
\end{table}

\section{Benchmarks and error measures}
\label{sec:benchmarks}

Four benchmarks are used, on the unit cube $\Omega = [0,1]^3$, ordered by the
extent to which their exact solutions depart from a signed-distance function.
The first two transport a smooth sphere rigidly and preserve its shape exactly;
the third transports a sphere carrying a narrow slot, whose reentrant edges make
the exact field non-differentiable; the fourth stretches the interface into a
thin sheet and returns it, so the field ceases to be a signed-distance function
over an extended region. The ordering is the one the results turn on.

\paragraph{Translating sphere (TR3D).}
A sphere of radius $0.15$ centred at $(0.5, 0.2, 0.5)$ is carried by the
uniform velocity $\mathbf{u} = (0, 0.1, 0)$ to $(0.5, 0.7, 0.5)$ at $T = 5$.

\paragraph{Rotating sphere (RO3D).}
A sphere of radius $0.15$ centred at $(0.5, 0.75, 0.5)$ is carried by
solid-body rotation about the vertical axis through $(0.5, 0.5)$,
$\mathbf{u} = (-(y - 0.5),\, x - 0.5,\, 0)$. One full revolution completes at
$T = 2\pi$, so the shape returns to its initial position and the error at
$t = T$ measures accumulated drift over a closed orbit.

\paragraph{Zalesak slotted sphere (ZD3D).}
The three-dimensional analogue of the slotted disc of \citet{Zalesak1979}: the
same rotation applied to a sphere of radius $0.15$ from which a rectangular
slot of half-extents $(0.025, 0.125, 0.10)$ centred at $(0.5, 0.725, 0.5)$ has
been removed. The slot has finite depth in $z$ and is therefore a blind pocket:
a section near the pole shows an unbroken sphere while a section through the
middle shows the notch, so three mutually orthogonal sections are required to
characterise the solid.

\paragraph{Reversed single vortex (RV3D).}
A sphere of radius $0.15$ centred at $(0.35, 0.35, 0.35)$ is deformed by the
three-dimensional field of \citet{Enright2002},
\begin{equation}
\begin{aligned}
    u &= \phantom{-}2\sin^2(\pi x)\sin(2\pi y)\sin(2\pi z)\cos(\pi\tau), \\
    v &= -\sin(2\pi x)\sin^2(\pi y)\sin(2\pi z)\cos(\pi\tau), \\
    w &= -\sin(2\pi x)\sin(2\pi y)\sin^2(\pi z)\cos(\pi\tau),
\end{aligned}
\end{equation}
with $T = 2$. The field is exactly divergence free, so the enclosed volume is
constant in time; the factor $\cos(\pi\tau)$ integrates to zero over the
interval, so the flow reverses at $\tau = 1/2$ and the sphere reforms at
$t = T$. This is the only benchmark in which the interface is stretched, and
hence the only one in which the advected field ceases to be a signed distance
function over an extended region.

\subsection{Reference solutions}

For the three rigid benchmarks the exact solution is available in closed form,
being the initial field composed with the inverse of the rigid motion. The
reversed vortex admits none, so the reference is constructed by integrating the
characteristics $\mathrm{d}\mathbf{X}/\mathrm{d}\tau = T\mathbf{u}$ backward
from each snapshot with a fourth-order Runge--Kutta scheme and evaluating
$\phi_0$ at the foot of each characteristic; since $\phi$ is advected as a
passive scalar this is exact up to the integration error. That error is bounded
by a self-consistency check at $t = T$, where the flow has reversed exactly and
the reference must reproduce $\phi_0$: with $200$ steps the discrepancy is
$6.3\times10^{-6}$ in the maximum norm and $1.2\times10^{-7}$ in the root mean
square, three to four orders of magnitude below the errors being measured.

\subsection{Error measures}
\label{sec:metrics}

All quantities are evaluated on a uniform $201^3$ grid ($\Delta x = 0.005$,
chosen so that the section planes used in the figures fall exactly on grid
points) at six equally spaced snapshots $t = 0, T/5, \ldots, T$, and reported
both per snapshot and as a snapshot average. Writing $e = \hat{\phi} - \phi$,

\begin{itemize}
    \item the \textbf{absolute error} $\lVert e\rVert_2 =
    \bigl(\overline{e^2}\bigr)^{1/2}$, which on a domain of unit measure
    coincides with the discrete $L_2$ norm;

    \item the \textbf{relative error}
    $\lVert e\rVert_2 / \lVert\phi\rVert_2$, expressed as a percentage, in the
    convention standard in the physics-informed literature;

    \item the \textbf{volume conservation error}
    $|V_{\mathrm{pred}} - V_{\mathrm{exact}}| / V_{\mathrm{exact}}$, with
    $V_{\mathrm{pred}}$ obtained by counting cells for which
    $\hat{\phi} < 0$. For the slotted sphere, which has no convenient
    closed-form volume, the reference is the counted volume of the exact field
    at $t = 0$, so that prediction and reference share the same estimator and
    its leading-order bias cancels;

    \item the \textbf{feature-restricted classification errors}, defined for
    benchmarks whose interface carries a thin geometric feature. Let
    $\Omega_{\mathrm{sol}}(t) = \{\mathbf{x} : \phi(\mathbf{x},t) < 0\}$ be the
    exact solid region and let $\Omega_{\mathrm{fea}}(t)$ be the region the
    feature removes from it, that is the set of points lying inside the
    unfeatured body but outside the true solid. For the slotted sphere
    $\Omega_{\mathrm{fea}}$ is the intersection of the slot with the sphere,
    obtained by transporting the unslotted sphere by the same rigid motion. We
    then report
    \begin{equation}
        F_{\mathrm{fill}}
        = \frac{\bigl|\{\mathbf{x}\in\Omega_{\mathrm{fea}} :
                        \phih(\mathbf{x},t) < 0\}\bigr|}
               {\bigl|\Omega_{\mathrm{fea}}\bigr|},
        \qquad
        F_{\mathrm{erode}}
        = \frac{\bigl|\{\mathbf{x}\in\Omega_{\mathrm{sol}} :
                        \phih(\mathbf{x},t) \ge 0\}\bigr|}
               {\bigl|\Omega_{\mathrm{sol}}\bigr|},
        \label{eq:feature}
    \end{equation}
    the fraction of the feature wrongly predicted solid and the fraction of the
    solid wrongly predicted void, both evaluated by cell counting on the same
    grid. Together they form a symmetric difference restricted to the feature.
    Unlike the volume error these are sign-based and cannot cancel: filling part
    of the feature while eroding the outer surface leaves the volume error small
    but raises both of these;

    \item the \textbf{eikonal deviation}
    $\overline{\bigl|\lVert\nabla\hat{\phi}\rVert - 1\bigr|}$, restricted to the
    band $|\phi| < 3\Delta x$ about the interface. The restriction excludes the
    gradient singularity at the centre of the sphere, where the exact signed
    distance function is not differentiable and a nonzero deviation is expected
    of any smooth approximation.
\end{itemize}

The cell-counting volume estimator has a resolution floor, which we quantify
rather than assume: applying it to the \emph{exact} field at each snapshot and
comparing against the reference gives a noise floor that is reported alongside
the volume errors and is below $0.3\%$ on every benchmark at this resolution.

\section{The fixed-weight baseline}
\label{sec:eik}

Before an adaptive scheme can be assessed it needs a target. This section
establishes, for each benchmark, the eikonal weight that a systematic search
selects and the sensitivity of the solution to that choice. These sweeps are
the ground truth against which SAW is measured in Section~\ref{sec:saw:results};
they are not themselves the contribution.

\subsection{Experimental protocol}
\label{sec:eik:protocol}

For each benchmark, $\weik$ was swept over six logarithmically spaced values
including $\weik = 0$ as a control that removes the regulariser entirely, each
trained with three random seeds (42, 43, 44), giving 18 independent runs per
benchmark. The seed controls both network initialisation and collocation
sampling, so the reported spread covers both sources of variability. Every run
used the full training budget of the reported configuration, so the ranking
transfers without extrapolation from truncated training. Component losses were
logged unweighted, so runs at different $\weik$ remain comparable term by term.

The selection rule was fixed before the sweeps were run: lowest seed-averaged
relative $L_2$ error, with ties broken by volume conservation, then by eikonal
deviation, then by the smaller weight. Two weights are treated as tied when the
gap between their means is smaller than the larger of their two seed standard
deviations. Eikonal deviation is deliberately demoted to a second-order
tiebreak: it is, up to the band restriction, the very quantity the swept term
minimises, so ranking by it would recover the sweep by construction and select
the largest weight irrespective of solution accuracy.

\subsection{Selected weights and sensitivity}
\label{sec:eik:selected}

Table~\ref{tab:fixed_sweeps} collects the outcome. The weight is consequential
on every benchmark: across the swept range the relative $L_2$ error varies by a
factor of up to $31$ and the volume conservation error by up to a factor of
$70$. The response is non-monotone with a well-defined interior optimum on
three of the four benchmarks; on the reversed vortex it is monotone, with the
minimum at the smallest weight tested.

\begin{table}[htbp]
\centering
\caption{Fixed-weight sweeps: the value selected on each benchmark, its
seed-averaged metrics, and the range spanned across the six weights examined.
Mean $\pm$ standard deviation over three seeds. The ZD3D grid is shifted one
decade below that of the sphere benchmarks and the RV3D grid one decade below
that again, since the appropriate weight falls with departure from the
signed-distance property.}
\label{tab:fixed_sweeps}
\begin{tabular}{lccccc}
\toprule
& & \multicolumn{2}{c}{at the selected weight}
& \multicolumn{2}{c}{range across the sweep} \\
\cmidrule(lr){3-4}\cmidrule(lr){5-6}
Benchmark & selected $\weik$ & rel. $L_2$ (\%) & vol. MAPE (\%)
          & rel. $L_2$ (\%) & vol. MAPE (\%) \\
\midrule
TR3D & $10^{-1}$ & $0.442 \pm 0.227$ & $1.207 \pm 0.240$ & $0.44$--$13.82$ & $0.81$--$20.41$ \\
RO3D & $10^{-1}$ & $0.297 \pm 0.205$ & $0.913 \pm 0.984$ & $0.30$--$1.64$  & $0.91$--$30.53$ \\
ZD3D$^{\dagger}$ & $10^{-3}$ & $0.516 \pm 0.105$ & $1.337 \pm 0.863$ & $0.52$--$2.41$ & $1.34$--$17.98$ \\
RV3D & $10^{-5}$ & $0.588 \pm 0.026$ & $1.173 \pm 0.639$ & $0.58$--$15.12$ & $1.17$--$82.59$ \\
\bottomrule
\end{tabular}

\vspace{2mm}
\begin{minipage}{0.92\textwidth}
\footnotesize $^{\dagger}$ ZD3D is reported at $N_f = 2\times10^{4}$,
$N_i = 10^{4}$. At the budget used for the other three benchmarks the slot is
not resolved: the volume error is already $6.28\%$ at $t=0$, before any
transport has occurred and at a time where the initial-condition term is
directly supervised, because roughly twelve of the $5\times10^{3}$
initial-condition points fall inside a slot occupying $2.36\times10^{-3}$ of the
domain. Doubling the budget reduces that error by an order of magnitude. The
same weight is selected at both budgets, so the selection is robust to the
sampling budget.
\end{minipage}
\end{table}

Three features of the sweeps are worth recording because they bear on what
follows.

\paragraph{The regulariser buys reproducibility as well as accuracy.}
At small weights the seed standard deviation is of the same order as the error
itself (on TR3D it reaches $7.8$ percentage points on a mean of $13.8$)
and at the selected weight it falls by more than an order of magnitude.
Differences below that spread are not resolvable from single runs. On TR3D the
selection rule declares a nominal tie between $10^{-1}$ and $10^{-2}$ for
exactly this reason: the gap between the means is smaller than the standard
deviation of the \emph{worse} candidate, which is itself large because one of
its three seeds produced an outlier of $15.66\%$ against $0.49\%$ and $1.34\%$.
The tie is an artefact of variance rather than a genuine equivalence, and the
criterion is reported as pre-registered rather than revised after inspection.

\paragraph{The second optimisation stage is not uniformly active.}
On RO3D the improvement attributable to L-BFGS was bimodal across the 18 runs:
in 13 it was of order $10^{-8}$ with a wall-clock time of $12.1$--$12.2$
minutes, and in 5 it was of order $10^{-4}$ with $16.4$--$17.5$ minutes. In the
latter group the strong-Wolfe line search located an admissible step from the
annealed Adam iterate; in the former it did not. Stratifying by engagement
leaves the selection intact (among the engaged runs the two at
$\weik = 10^{-1}$ gave $0.143\%$ and $0.160\%$ against $0.197$--$0.426\%$
elsewhere, and among the dormant runs the single run at $10^{-1}$ gave the
lowest value in that group), but the reported standard deviations are
inflated by it.

\paragraph{A large weight is harmful where the exact solution is not an SDF.}
On TR3D volume conservation is best at the largest weight tested; on RO3D it is
best at the selected weight and degrades by a factor of five at $\weik = 1$. On
ZD3D it is minimised at the selected weight and then degrades sharply, from
$1.34\%$ at $10^{-3}$ to $17.98\%$ at $10^{-1}$, and on RV3D the relative $L_2$ error rises by an order
of magnitude between $10^{-5}$ and $10^{-4}$ and the volume error reaches
$82.6\%$ at $10^{-3}$. The weight that is optimal on both smooth spheres is
among the worst settings on the other two.

\subsection{What the selected weights have in common}
\label{sec:eik:growth}

The selected values span four decades and they order by how far the exact
solution departs from a signed-distance function.

\begin{itemize}
    \item \textbf{No departure.} TR3D and RO3D transport a smooth sphere by a
    rigid motion, so $\lVert\nabla\phi\rVert = 1$ holds exactly for all time,
    the eikonal constraint is everywhere consistent with the governing equation,
    and both select $\weik = 10^{-1}$.
    \item \textbf{Local departure.} ZD3D is also a rigid motion, but its
    constructive-solid-geometry initial field
    $\phi_0 = \max(\phi_{\mathrm{sphere}}, -\phi_{\mathrm{slot}})$ returns the
    distance to the nearer face rather than to the edge near the reentrant edges
    of the slot. The eikonal property holds almost everywhere, but the maximum
    introduces gradient discontinuities along those edges, and a smooth network
    cannot satisfy $\lVert\nabla\phi\rVert = 1$ in a neighbourhood of a crease it
    cannot represent. The selected
    weight falls two decades to $10^{-3}$.
    \item \textbf{Global departure.} RV3D stretches the interface, so the
    advected field ceases to be a signed-distance function throughout the
    deformed region. The selected weight falls two decades further, to
    $10^{-5}$.
\end{itemize}

The eikonal deviation measured within a single RV3D run makes the mechanism
explicit: at the selected weight it rises from $2.06\times10^{-2}$ at $t=0$ to
$1.23$ near maximum deformation and falls back to $4.67\times10^{-2}$ at $t=T$,
a variation of nearly two orders of magnitude. The departure is therefore not a
defect of the approximation but a property of the exact solution being
approximated: it appears as the interface stretches and disappears as it
reforms. This is the observation SAW is built on.

The same ordering was reported for the two-dimensional counterparts of these
benchmarks~\citep{Khan2026}; the present sweeps establish that it survives the
transition to three dimensions, where the collocation domain gains a dimension
and the sampling density per dimension falls accordingly. The individual values
transfer less reliably: two of the
four carry over unchanged, while the translating sphere moves from $1$ to
$10^{-1}$ (at which the two-dimensional value is three times worse) and
the reversed vortex from $10^{-4}$ to $10^{-5}$.

The practical implication is that a weight established in two dimensions cannot
be assumed to remain appropriate in three, but neither can it be assumed to
change: transfer is benchmark-dependent and has to be checked. The ordering is
predictive of the decade; it is not predictive of the value. Locating the value
costs eighteen training runs per benchmark, between four and thirteen hours on a
single T4 GPU, and the cost is incurred again for every new problem.

\subsection{Performance at the selected weights}
\label{sec:eik:reported}

Table~\ref{tab:reported} gives the metrics for the single run at each selected
weight that supplies the figures of this section. In each case the run
reproduces the
corresponding sweep entry to the quoted precision, since the seed and every
other setting are unchanged.

\begin{table}[htbp]
\centering
\caption{Metrics at each benchmark's selected weight, seed 42, evaluated on the
$201^3$ grid and averaged over the six snapshots. Timings are for a single
NVIDIA T4.}
\label{tab:reported}
\begin{tabular}{lccccc}
\toprule
Benchmark & abs. $L_2$ & rel. $L_2$ (\%) & volume MAPE (\%)
          & eikonal dev. & time (min) \\
\midrule
TR3D & $4.73\times10^{-4}$ & $0.122$ & $1.143$ & $1.31\times10^{-2}$ & $11.8$ \\
RO3D & $6.76\times10^{-4}$ & $0.160$ & $0.117$ & $2.08\times10^{-2}$ & $15.6$ \\
ZD3D$^{\dagger}$ & $2.03\times10^{-3}$ & $0.480$ & $0.328$ & $1.02\times10^{-1}$ & $24.9$ \\
RV3D & $2.52\times10^{-3}$ & $0.591$ & $0.564$ & $5.97\times10^{-1}$ & $23.7$ \\
\bottomrule
\end{tabular}

\vspace{2mm}
\begin{minipage}{0.92\textwidth}
\footnotesize $^{\dagger}$ At the doubled sampling budget, for the reason given
in Table~\ref{tab:fixed_sweeps}.
\end{minipage}
\end{table}

The benchmarks trade off differently between the two error measures.
Translation attains the lower field error but a volume error of $1.14\%$ which
is nearly constant across snapshots and therefore represents a systematic
sub-cell offset in the interface radius rather than progressive loss: the
implied radius error is $0.38\%$, approximately one ninth of a grid cell at
$\Delta x = 0.005$. Rotation attains a slightly larger field error but conserves
volume to $0.117\%$, close to the resolution floor of the cell-counting
estimator on this grid. The two benchmarks whose exact solutions are not
signed-distance functions carry field errors roughly four times larger than
TR3D while still conserving volume to better than $0.6\%$; their eikonal
deviations are larger by factors of eight and forty-five respectively, which
reflects the geometry of the exact solution rather than a deficiency of the
approximation.

\begin{figure}[htbp]
\centering
\includegraphics[width=\textwidth]{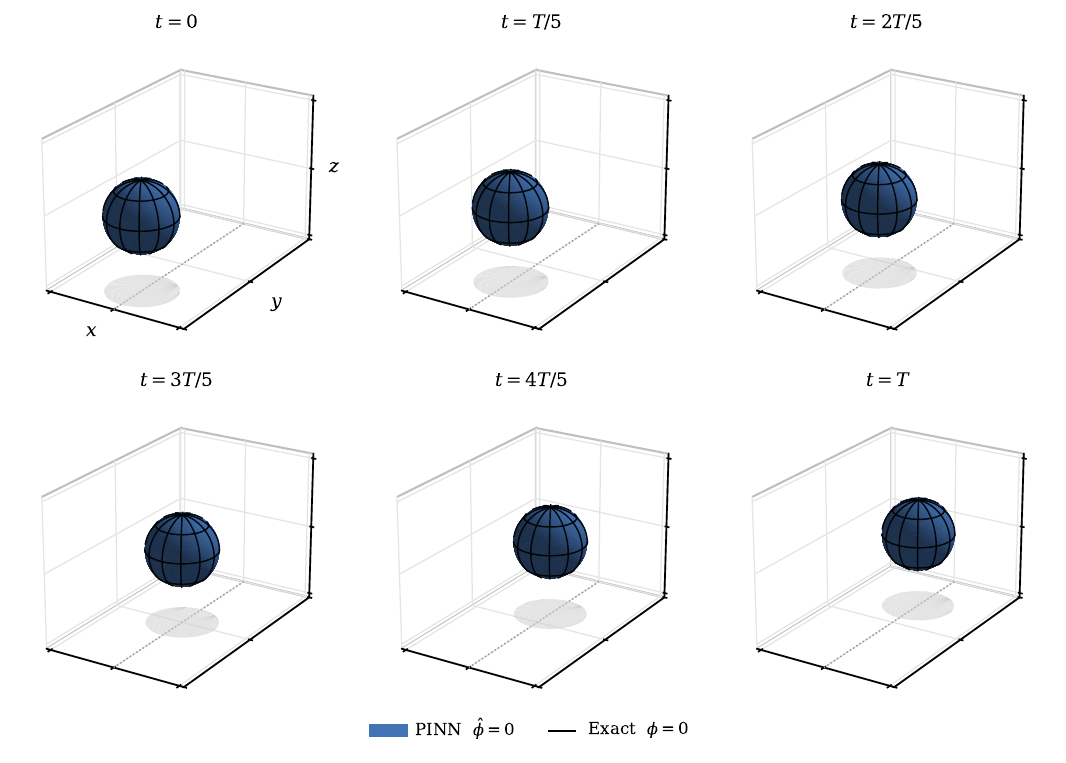}
\caption{Translating sphere at $\weik = 10^{-1}$. Shaded surface: predicted zero
level set $\phih = 0$, extracted by marching cubes~\citep{Lorensen1987} on the
$201^3$ evaluation
grid. Black wireframe: exact interface. The dotted line and grey disc on the
lower plane mark the centre trajectory and its projection. Axes are cropped to
$x, z \in [0.18, 0.82]$ and $y \in [0,1]$.}
\label{fig:tr3d_interface}
\end{figure}

\begin{figure}[htbp]
\centering
\includegraphics[width=\textwidth]{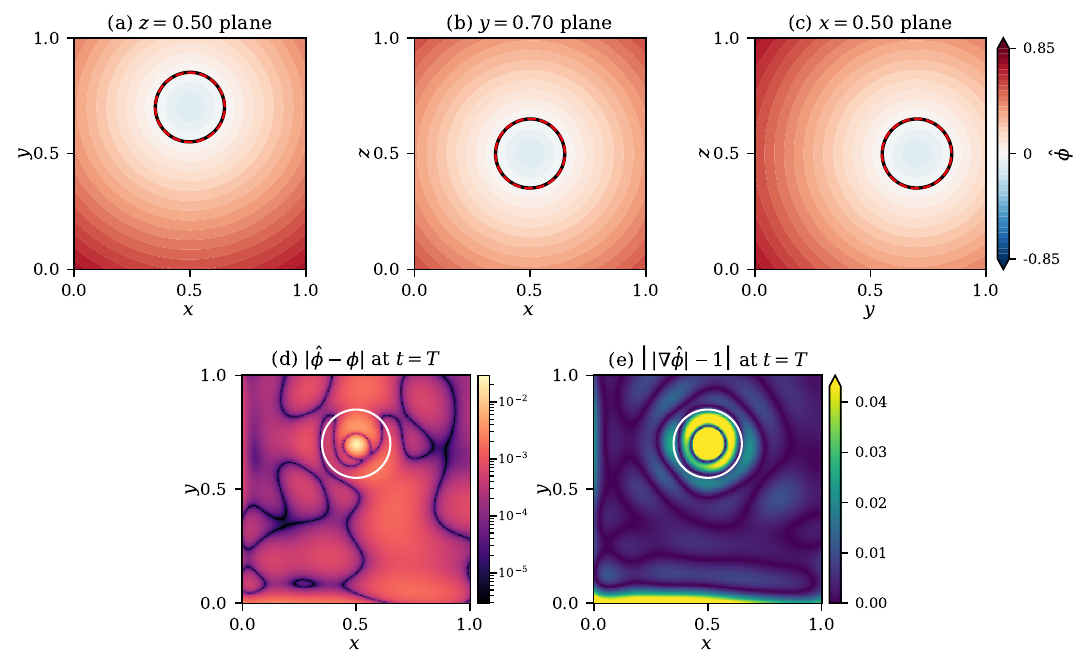}
\caption{Verification for the translating sphere at $\weik = 10^{-1}$.
(a)--(c) Three mutually orthogonal sections through the interface centre at
$t=T$, with the predicted zero contour (solid black) and the exact one (dashed
red) superposed on the predicted field; three circular cross-sections
distinguish a sphere from a cylinder, which a single section cannot.
(d) Pointwise error at $t=T$ on a logarithmic scale, the peak coinciding with
the gradient singularity at the sphere centre. (e) Eikonal deviation,
colour-scaled to the $99$th percentile within $\lvert\phi\rvert < 5\Delta x$.}
\label{fig:tr3d_verification}
\end{figure}

\begin{figure}[htbp]
\centering
\includegraphics[width=\textwidth]{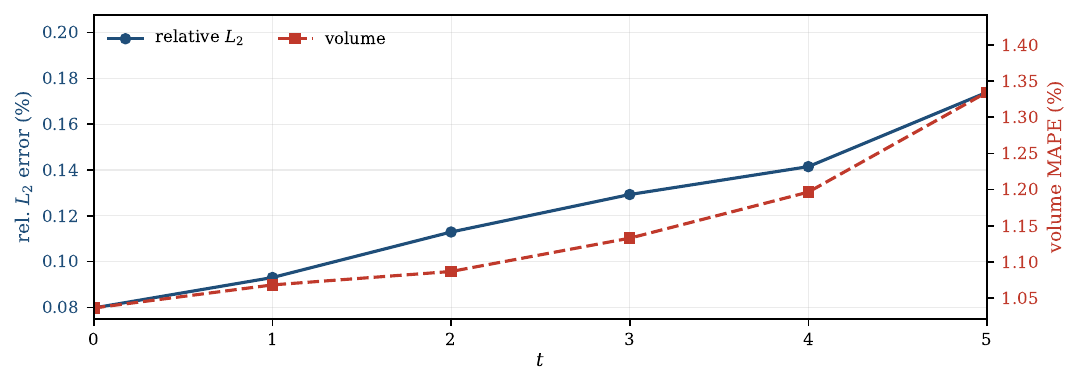}
\caption{Growth of relative $L_2$ error and volume MAPE for the translating
sphere at $\weik = 10^{-1}$. Both curves are close to flat: the field error
grows by a factor of $3.7$ over the interval, against a factor of $79$ with the
regulariser switched off.}
\label{fig:tr3d_growth}
\end{figure}

\begin{figure}[htbp]
\centering
\includegraphics[width=\textwidth]{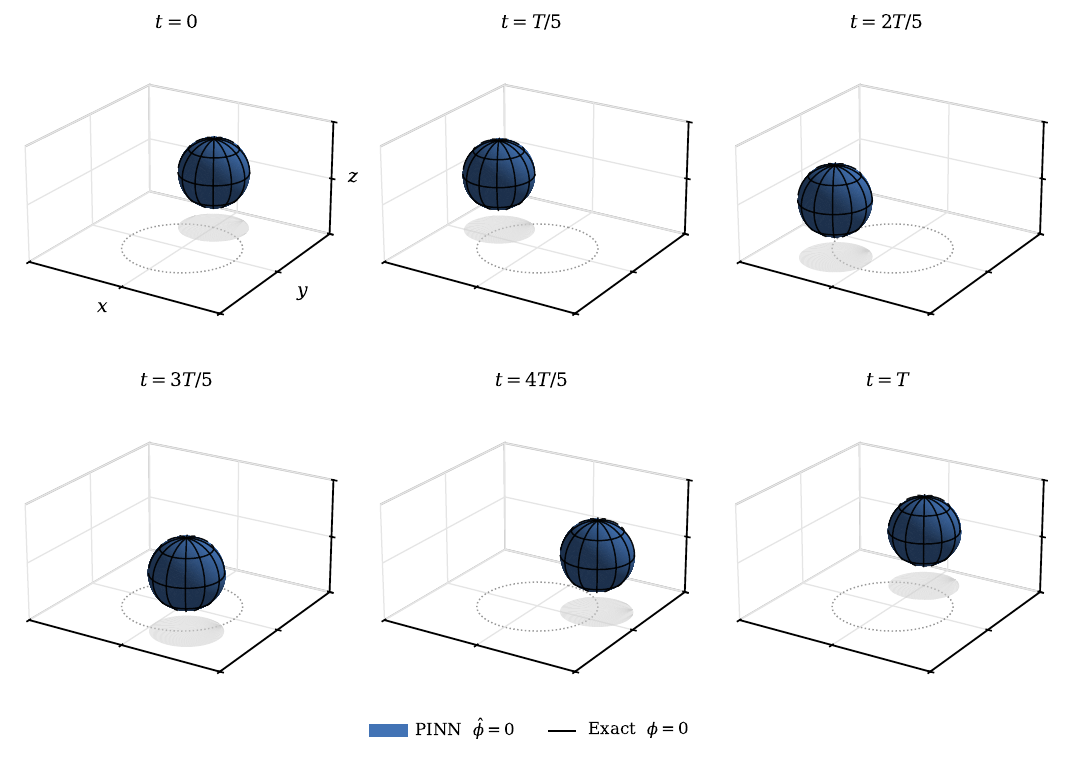}
\caption{Rotating sphere at $\weik = 10^{-1}$. The dotted circle on the lower
plane is the orbit of the sphere centre. One full revolution completes at $t=T$,
so the final panel returns to the initial configuration and its error measures
accumulated drift over a closed orbit.}
\label{fig:ro3d_interface}
\end{figure}

\begin{figure}[htbp]
\centering
\includegraphics[width=\textwidth]{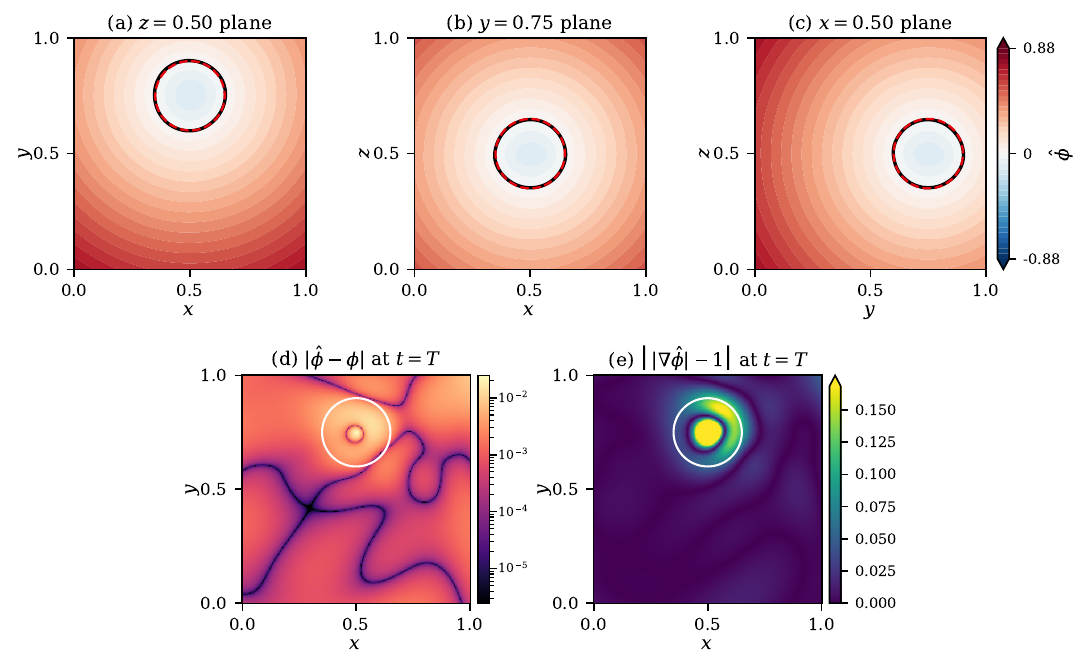}
\caption{Verification for the rotating sphere at $\weik = 10^{-1}$. Panels as in
Figure~\ref{fig:tr3d_verification}.}
\label{fig:ro3d_verification}
\end{figure}

\begin{figure}[htbp]
\centering
\includegraphics[width=\textwidth]{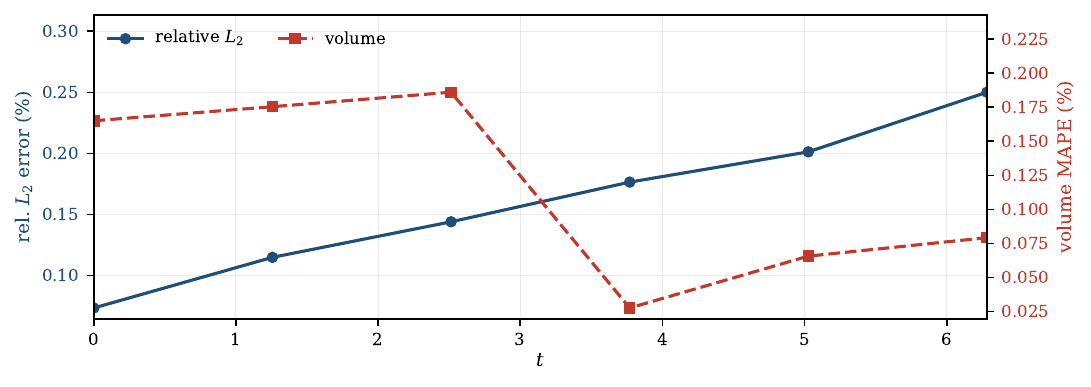}
\caption{Growth of relative $L_2$ error and volume MAPE for the rotating sphere
at $\weik = 10^{-1}$. The final point measures accumulated drift over a closed
orbit.}
\label{fig:ro3d_growth}
\end{figure}

Figures~\ref{fig:tr3d_interface}--\ref{fig:ro3d_growth} show the fixed-weight
solutions on the two benchmarks whose exact solutions are signed-distance
functions throughout, where the fixed weight is also the reported model. For
each the first figure gives the predicted zero level set at six snapshots with
the exact interface overlaid, the second gives three mutually orthogonal
sections at $t=T$ with the pointwise error and eikonal deviation, and the third
gives the growth of both error measures over the integration window. The
corresponding figures for the slotted sphere and the reversed vortex appear in
Section~\ref{sec:saw:results}, where SAW supplies the reported model.

\section{SDF-Aware Weighting: results}
\label{sec:saw:results}

SAW was applied to all four benchmarks with a single configuration
($\beta = 0.999$, $q = 0.95$) and three seeds per setting, matching the
protocol of Section~\ref{sec:eik:protocol} in every other respect. The gate was
additionally disabled ($q = 1$), which reduces SAW to plain gradient-norm
balancing; that column is an ablation rather than a separate experiment, and it
is run under the same protocol so that its outcome is comparable.

\subsection{SAW against the fixed-weight baseline}
\label{sec:saw:comparison}

Table~\ref{tab:saw} reports the outcome. The result divides cleanly along the
line that Section~\ref{sec:eik:growth} identifies. On the two benchmarks whose
exact solutions are not smooth signed-distance functions, SAW is competitive with
or better than the weight that eighteen runs of search selected: on the slotted
sphere it attains $0.285\%$ against $0.516\%$, a factor of $1.8$, and lower than
\emph{any} of the six weights that sweep examined. On the reversed vortex it
attains $0.868\%$ against $0.588\%$, within a factor of $1.5$ of a sweep it did
not have to perform.

On the two smooth rigid benchmarks it is worse. This is the expected outcome and
is discussed in Section~\ref{sec:saw:controls}.

\begin{table}[htbp]
\centering
\caption{SAW ($\beta = 0.999$, $q = 0.95$) against the fixed-weight baseline of
Table~\ref{tab:fixed_sweeps}, and against SAW with the gate disabled. Mean
$\pm$ standard deviation over three seeds, except where noted. The final column
is the eikonal weight SAW converged to, seed-averaged.}
\label{tab:saw}
\begin{tabular}{llccc}
\toprule
Benchmark & configuration & rel. $L_2$ (\%) & vol. MAPE (\%) & $\weik$ \\
\midrule
\multirow{3}{*}{TR3D}
 & fixed sweep, $\weik = 10^{-1}$ & $\mathbf{0.442 \pm 0.227}$ & $\mathbf{1.207 \pm 0.240}$ & $10^{-1}$ \\
 & SAW, $q = 0.95$                & $1.035 \pm 0.539$ & $1.811 \pm 0.526$ & $5.11\times10^{-2}$ \\
 & SAW, gate off                  & $0.449 \pm 0.193^{\ast}$ & $0.191 \pm 0.024^{\ast}$ & $2.14\times10^{-1}$ \\
\midrule
\multirow{3}{*}{RO3D}
 & fixed sweep, $\weik = 10^{-1}$ & $\mathbf{0.297 \pm 0.205}$ & $\mathbf{0.913 \pm 0.984}$ & $10^{-1}$ \\
 & SAW, $q = 0.95$                & $0.816 \pm 0.032$ & $19.045 \pm 1.251$ & $1.79\times10^{-1}$ \\
 & SAW, gate off                  & $1.593 \pm 0.096$ & $5.762 \pm 1.176$  & $1.00$ \\
\midrule
\multirow{3}{*}{ZD3D}
 & fixed sweep, $\weik = 10^{-3}$ & $0.516 \pm 0.105$ & $1.337 \pm 0.863$ & $10^{-3}$ \\
 & SAW, $q = 0.95$                & $\mathbf{0.285 \pm 0.155}$ & $\mathbf{1.561 \pm 1.167}$ & $2.25\times10^{-3}$ \\
 & SAW, gate off                  & $1.057 \pm 0.015$ & $17.573 \pm 0.483$ & $7.03\times10^{-1}$ \\
\midrule
\multirow{3}{*}{RV3D}
 & fixed sweep, $\weik = 10^{-5}$ & $\mathbf{0.588 \pm 0.026}$ & $\mathbf{1.173 \pm 0.639}$ & $10^{-5}$ \\
 & SAW, $q = 0.95$                & $0.868 \pm 0.036$ & $1.218 \pm 0.439$ & $8.25\times10^{-5}$ \\
 & SAW, gate off                  & $14.967 \pm 0.556$ & $76.100 \pm 8.974$ & $1.00$ \\
\bottomrule
\end{tabular}

\vspace{2mm}
\begin{minipage}{0.92\textwidth}
\footnotesize $^{\ast}$ Two seeds. The third diverged during the L-BFGS stage
after $1002$ closure evaluations. The adaptive weight rose from $0.600$ to
$0.815$ across the L-BFGS stage and was flat over the final hundred evaluations
before the failure, which was a line-search step producing non-finite
parameters. This is the ungated
configuration on a control benchmark, that is, SAW applied outside its stated
scope.
\end{minipage}
\end{table}

\subsection{Ablation: the role of the gate}
\label{sec:saw:ablation}

The two components of SAW are not independent, and the $q = 1$ column shows why.
With the gate disabled the weight fails to descend to the value the sweep
identifies. On RO3D and RV3D it remains at exactly $1.00$, its initial value and
its clip, for the whole of training; on ZD3D it reaches only $0.70$, still two
and a half decades above the swept optimum of $10^{-3}$. Only on TR3D, where the
exact solution is a signed-distance function everywhere and the eikonal residual
does converge, does the ungated ratio settle near the correct value. The
mechanism is visible in the loss trace on the two benchmarks where the property
fails; on RO3D the weight also fails to descend, for reasons the present runs do
not isolate. Where the exact solution violates the signed-distance property, the
ungated eikonal loss
is dominated by the points at which that violation occurs, and those points do
not converge; $\Leik$ therefore stays small in the mean while
$\lVert\nabla_{\!\theta}\Leik\rVert$ stays small with it, the ratio
$(g_{\mathrm{pde}} + g_{\mathrm{ic}})/g_{\mathrm{eik}}$ stays large, and the
descent stalls far above the value the sweep identifies. Removing those points is what allows the remaining
residual to carry information the ratio can act on.

The consequences are severe on the benchmark where the departure is global. On
the reversed vortex, disabling the gate degrades the relative $L_2$ error by a
factor of $17$ and the volume error by a factor of $62$, the latter reaching
$76\%$; the zero level set is effectively lost. On the slotted sphere the
factors are $3.7$ and $11$. Ungated gradient-norm balancing is worse on both
benchmarks than the weight the sweep selected in Table~\ref{tab:fixed_sweeps}.

This is the central empirical claim of the paper: adaptive loss balancing, in
the form in which it is normally applied, fails on this problem class, and it
fails for a reason that is structural rather than a matter of tuning.

\subsection{Weight recovery across benchmarks}
\label{sec:saw:weight}

Table~\ref{tab:saw_weight} compares the weight SAW converges to against the
value the sweep selected. With one fixed configuration the adaptive weight lands
within an order of magnitude of the swept optimum on every benchmark, spanning
four decades from $10^{-1}$ to $10^{-5}$. The agreement is closest where the
signed-distance property fails locally (a factor of $2.3$ on the slotted sphere)
and loosest where it fails globally (a factor of $8.3$ on the reversed vortex);
the ratio balances gradient norms rather than minimising the reported error, so
exact agreement is not expected. What is reproduced exactly is the ordering.

\begin{table}[htbp]
\centering
\caption{The weight SAW converges to ($\beta = 0.999$, $q = 0.95$, mean over
three seeds) against the value selected by the eighteen-run sweep.}
\label{tab:saw_weight}
\begin{tabular}{lccc}
\toprule
Benchmark & swept optimum & SAW converged to & ratio \\
\midrule
TR3D & $10^{-1}$ & $5.11\times10^{-2}$ & $0.5$ \\
RO3D & $10^{-1}$ & $1.79\times10^{-1}$ & $1.8$ \\
ZD3D & $10^{-3}$ & $2.25\times10^{-3}$ & $2.3$ \\
RV3D & $10^{-5}$ & $8.25\times10^{-5}$ & $8.3$ \\
\bottomrule
\end{tabular}
\end{table}

The weight is not on a schedule. It begins at its clip and remains there while
the eikonal loss is small, then descends as the interface deforms and the
eikonal residual grows. On the reversed vortex it falls from $1$ to
$8\times10^{-5}$ between iterations $6000$ and $20000$; on the slotted sphere it
holds near $0.2$--$0.3$ over a long plateau before descending to
$2\times10^{-3}$. The descent is driven by the growth of
$\lVert\nabla_{\!\theta}\Leik\rVert$, which is itself a consequence of the exact
solution moving away from the signed-distance property.

\begin{figure}[htbp]
\centering
\includegraphics[width=\textwidth]{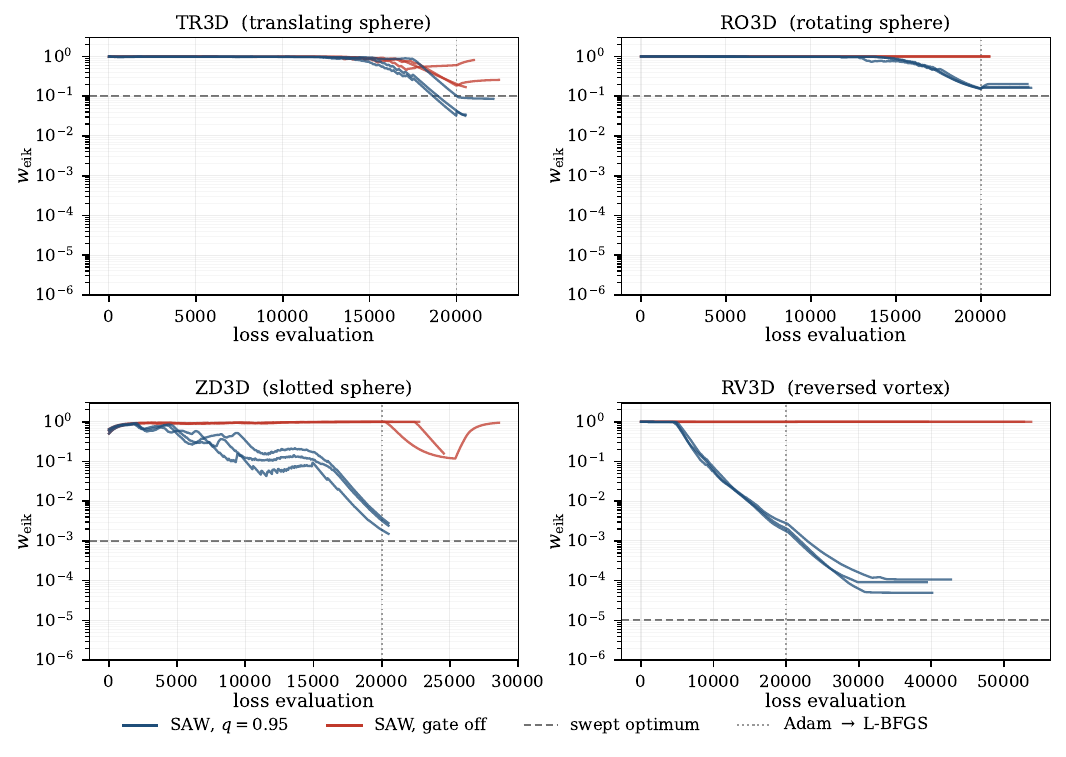}
\caption{The adaptive weight over training, three seeds per configuration.
The gated runs descend from the clip towards the value the eighteen-run sweep
selected (dashed); with the gate disabled the weight does not descend. The
horizontal axis counts loss evaluations rather than iterations, since the
L-BFGS line search calls the closure several times per step.}
\label{fig:saw_trajectory}
\end{figure}

\subsection{Feature-restricted error measures}
\label{sec:saw:feature}

The slotted sphere is the one benchmark carrying a thin geometric feature, and
Table~\ref{tab:saw_feature} reports the measures of
Equation~\eqref{eq:feature} for it. They make a distinction the relative $L_2$
error does not.

\begin{table}[htbp]
\centering
\caption{Feature-restricted classification errors on the slotted sphere, mean
$\pm$ standard deviation over three seeds. $F_{\mathrm{fill}}$ is the fraction
of the slot wrongly predicted solid, $F_{\mathrm{erode}}$ the fraction of the
solid wrongly predicted void.}
\label{tab:saw_feature}
\begin{tabular}{lccccc}
\toprule
& \multicolumn{3}{c}{$F_{\mathrm{fill}}$ (\%)} & & \\
\cmidrule(lr){2-4}
Configuration & at $t=0$ & at $t=T$ & mean
              & mean $F_{\mathrm{erode}}$ (\%) & rel. $L_2$ (\%) \\
\midrule
SAW, $q = 0.95$ & $12.5$ & $24.8$  & $19.7 \pm 5.0$ & $3.77 \pm 0.48$ & $0.285$ \\
SAW, gate off   & $55.1$ & $99.9$ & $92.4 \pm 2.6$ & $1.26 \pm 0.08$ & $1.057$ \\
\bottomrule
\end{tabular}
\end{table}

With the gate disabled the slot is all but entirely filled by the end of the
integration: $F_{\mathrm{fill}} = 99.9\%$ at $t = T$, meaning all but a
fraction of a percent of the removed region is wrongly classified as solid. The relative $L_2$ error for that
solution is $1.057\%$. Substituting a sphere with no slot at all (a field
that never represented the feature) gives $1.183\%$ on the same grid. The two
cases are therefore separated by about one tenth of a percentage point in the
measure conventionally reported, and by the full range of
$F_{\mathrm{fill}}$.

The two feature measures also move in opposite directions, which is why both are
needed. Disabling the gate raises $F_{\mathrm{fill}}$ from $19.7\%$ to $92.4\%$
but \emph{lowers} $F_{\mathrm{erode}}$ from $3.77\%$ to $1.26\%$: a solution
that fills the slot has less solid left to erode. A signed volume error, in
which the two contributions subtract, would understate the damage for exactly
this reason. Sign-based measures restricted to the feature cannot cancel in that
way.

\subsection{Negative controls}
\label{sec:saw:controls}

On TR3D and RO3D the exact solution is a signed-distance function everywhere and
for all time. There are no points at which a large eikonal residual indicates
legitimate departure; a large residual there is network error, and it is exactly
the supervision the eikonal term exists to provide. Gating the upper tail
discards it. SAW is correspondingly worse than the fixed weight on both: on TR3D
by a factor of $2.3$ in the field norm, and on RO3D by $2.7$ in the field norm
and $21$ in volume conservation.

We report this rather than restricting the paper to the benchmarks where the
method succeeds. A scheme that improved every case would be evidence of a
generic regularisation effect (gating a fraction of any loss term tends to
help a little) rather than of the mechanism claimed here. That SAW helps
where the signed-distance property fails and hurts where it holds is what
distinguishes the two explanations.

\begin{figure}[htbp]
\centering
\includegraphics[width=\textwidth]{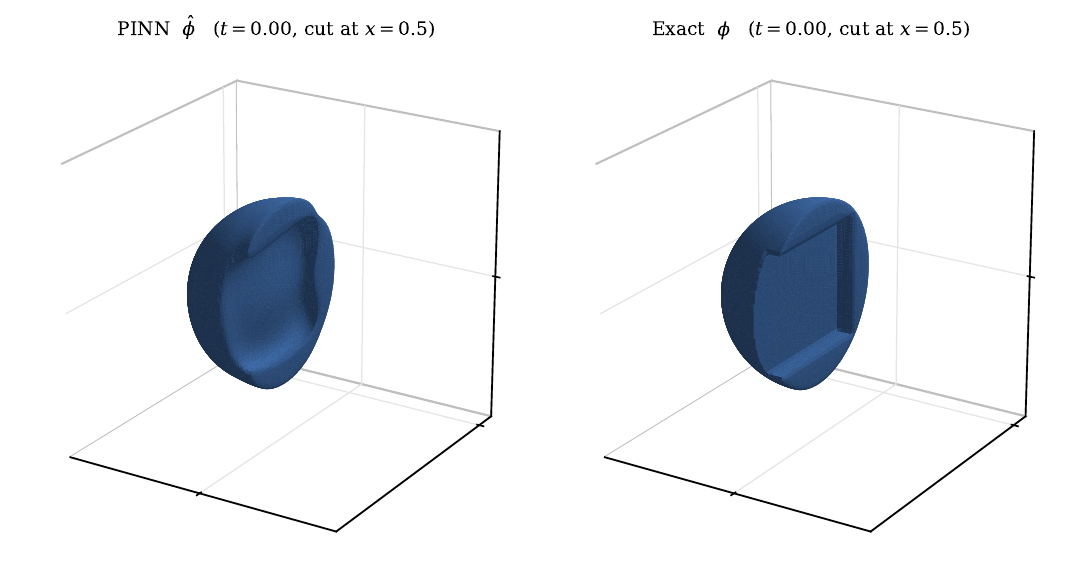}
\caption{Half-space cutaway at $x = 0.5$ through the slot, predicted (left) and
exact (right), at $t=0$, under SAW ($\beta = 0.999$, $q = 0.95$). The slot has
finite depth in $z$ and is therefore a
blind pocket: a section near the pole shows an unbroken sphere while a section
through the middle shows the notch. The mouth is a narrow opening near
$z = 0.5$, so from outside the shape reads as an almost intact sphere and this
view is required to establish the geometry.}
\label{fig:zd3d_cutaway}
\end{figure}

\begin{figure}[htbp]
\centering
\includegraphics[width=\textwidth]{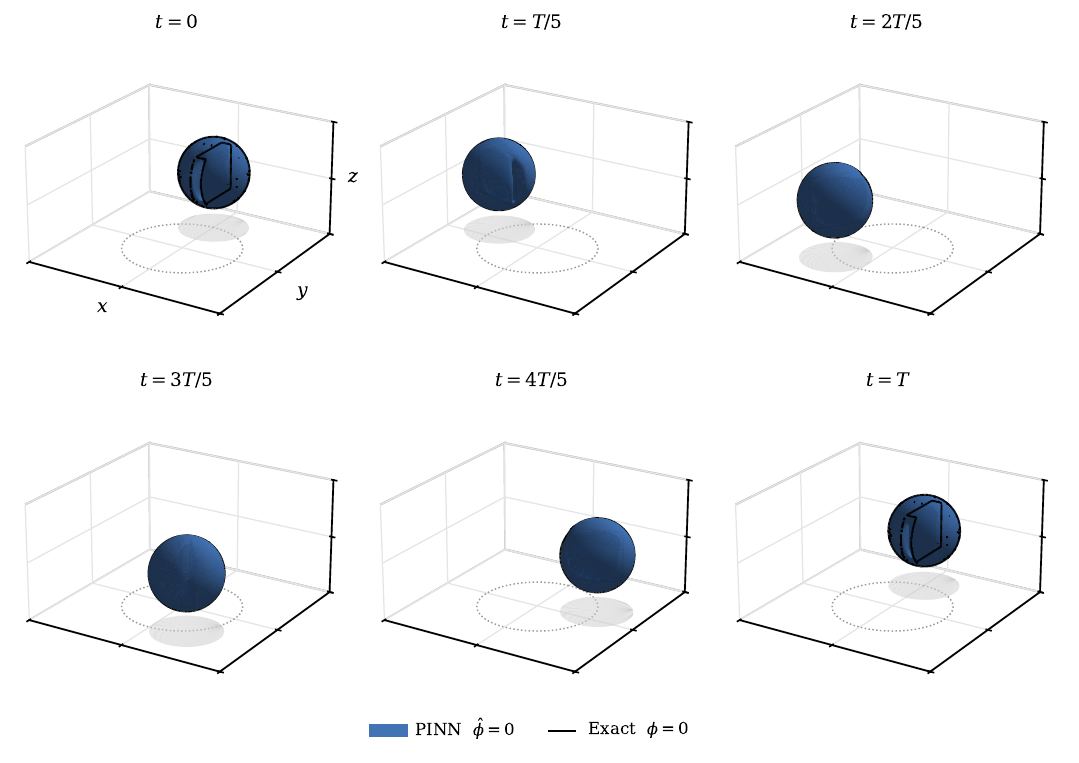}
\caption{Zalesak slotted sphere under SAW ($\beta = 0.999$, $q = 0.95$). Black
outline: silhouette
of the exact interface, obtained by culling back-facing triangles of the exact
isosurface against the view direction; the shape is non-convex, so the analytic
wireframe used in Figures~\ref{fig:tr3d_interface} and~\ref{fig:ro3d_interface}
does not apply.}
\label{fig:zd3d_interface}
\end{figure}

\begin{figure}[htbp]
\centering
\includegraphics[width=\textwidth]{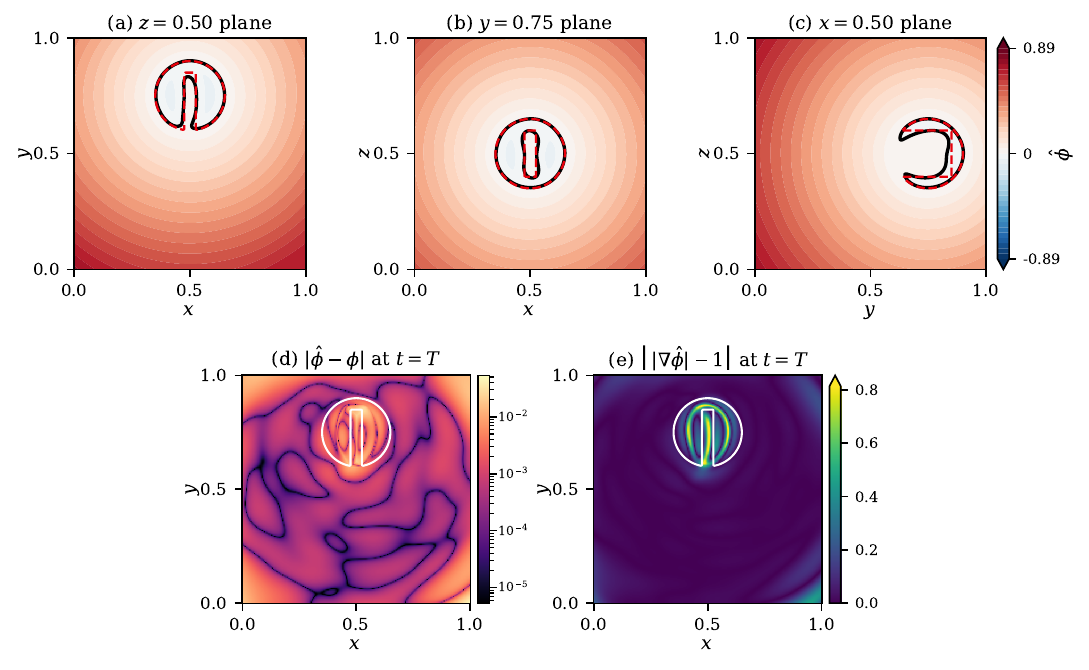}
\caption{Verification for the Zalesak slotted sphere under SAW
($\beta = 0.999$, $q = 0.95$).
Panels as in Figure~\ref{fig:tr3d_verification}. In (e) the elevated eikonal
deviation near the slot edges is expected: the exact field is non-differentiable
there and a smooth network cannot reproduce the crease, so a nonzero deviation is
not purely network error.
These are the points the residual-quantile gate excludes.}
\label{fig:zd3d_verification}
\end{figure}

\begin{figure}[htbp]
\centering
\includegraphics[width=\textwidth]{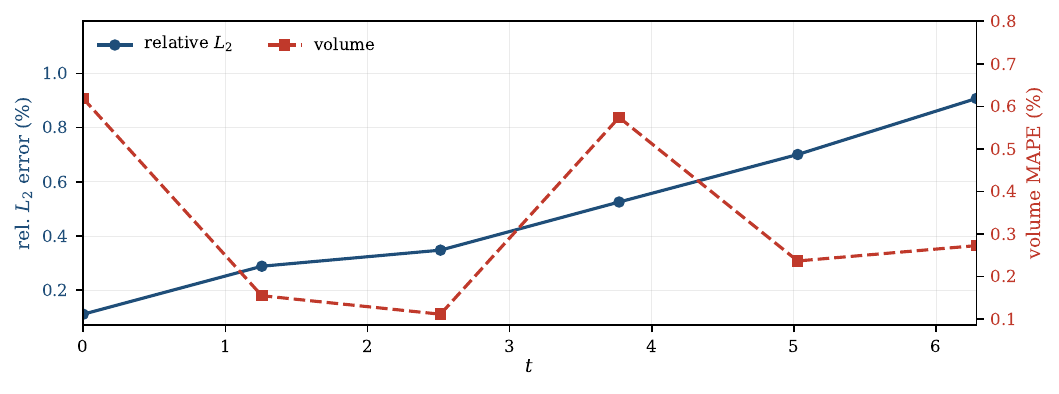}
\caption{Growth of relative $L_2$ error and volume MAPE for the Zalesak slotted
sphere under SAW ($\beta = 0.999$, $q = 0.95$). The two measures are not in
step: the volume error
is non-monotone while the field error grows steadily.}
\label{fig:zd3d_growth}
\end{figure}

\begin{figure}[htbp]
\centering
\includegraphics[width=\textwidth]{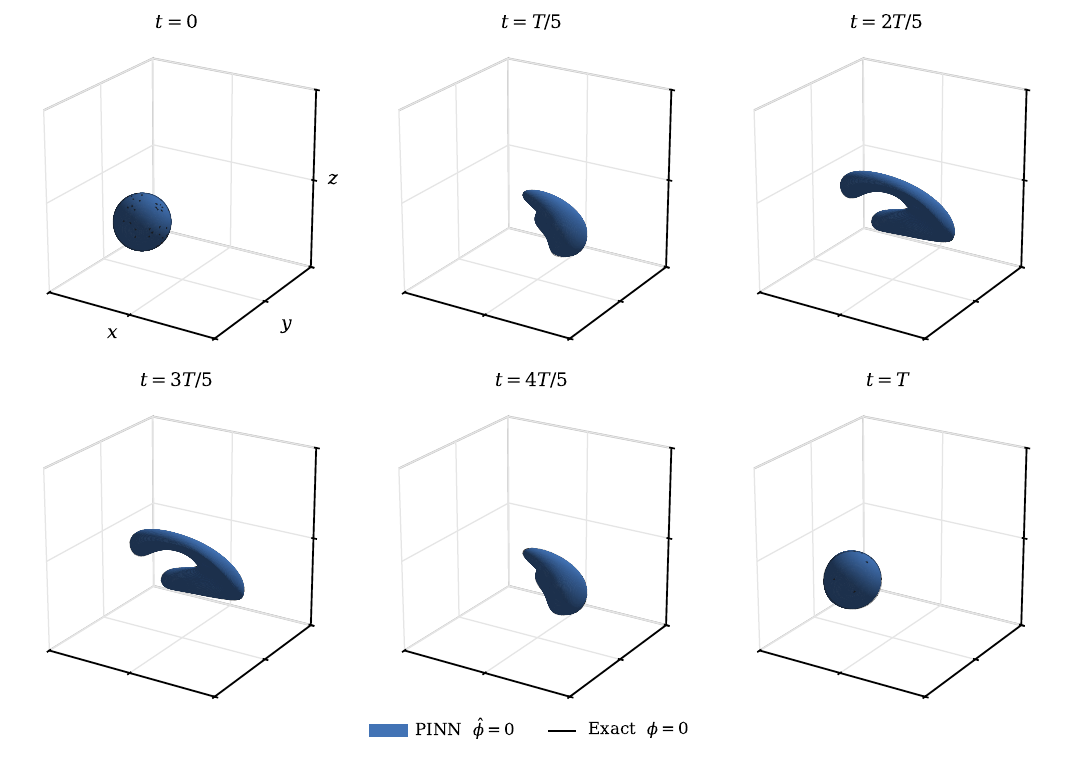}
\caption{Reversed single vortex under SAW ($\beta = 0.999$, $q = 0.95$). Black
outline: silhouette
of the exact interface obtained from the backtraced reference. The sphere is
stretched into a folded sheet, reaching maximum deformation at $t = T/2$, and
reforms at $t=T$.}
\label{fig:rv3d_interface}
\end{figure}

\begin{figure}[htbp]
\centering
\includegraphics[width=\textwidth]{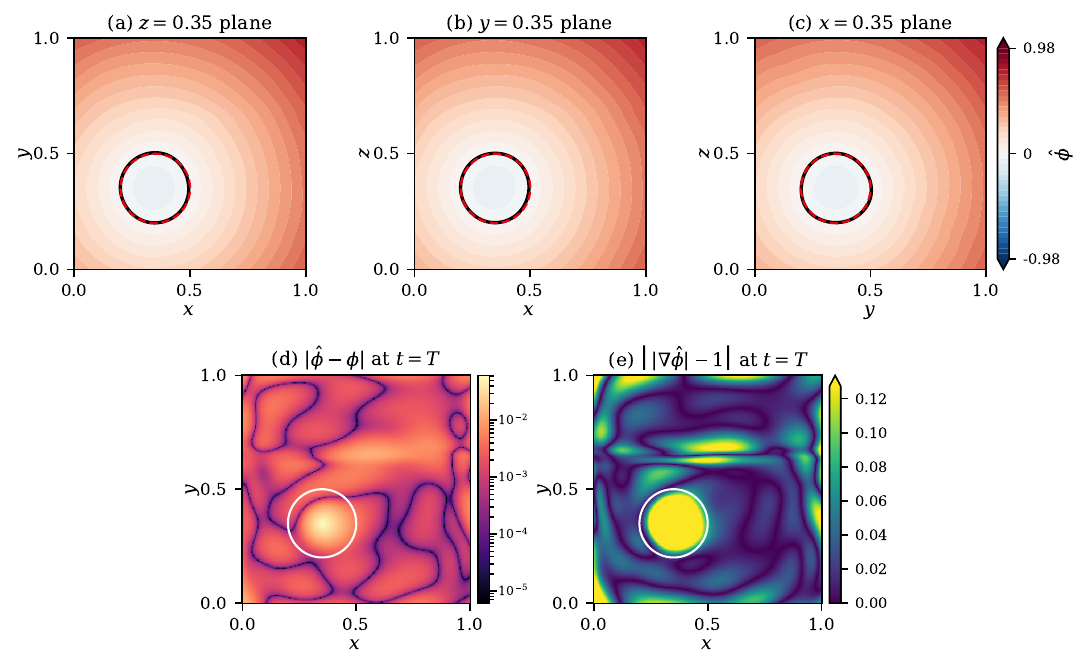}
\caption{Verification for the reversed single vortex under SAW
($\beta = 0.999$, $q = 0.95$).
Panels as in Figure~\ref{fig:tr3d_verification}, with sections taken through the
centre of the reformed sphere at $t=T$.}
\label{fig:rv3d_verification}
\end{figure}

\begin{figure}[htbp]
\centering
\includegraphics[width=\textwidth]{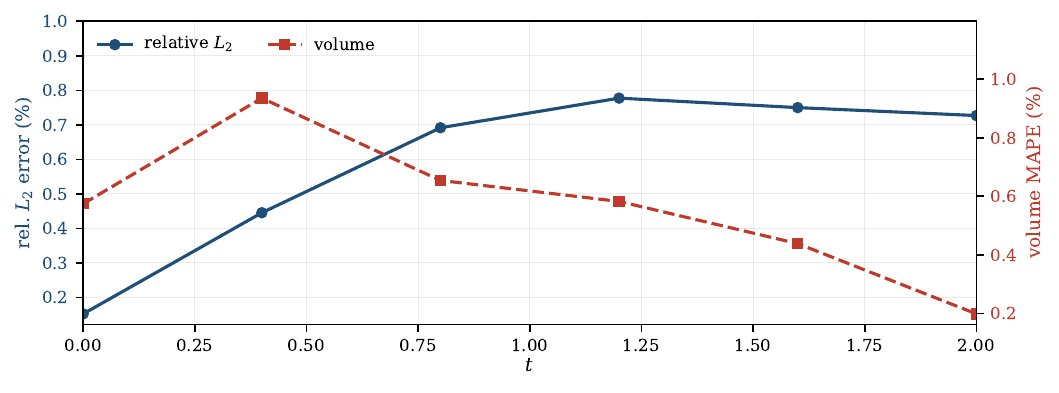}
\caption{Growth of relative $L_2$ error and volume MAPE for the reversed single
vortex under SAW ($\beta = 0.999$, $q = 0.95$). The interface reaches maximum
deformation near
$t = T/2$ and reforms at $t=T$.}
\label{fig:rv3d_growth}
\end{figure}

The controls also bound the applicability of the method. SAW should be applied
when the exact solution is known or expected to depart from a signed-distance
function: when the initial geometry is constructed by boolean operations, or
when the velocity field is not a rigid motion. Both conditions are properties of
the problem specification and can be checked before any training is performed.
Figures~\ref{fig:zd3d_cutaway}--\ref{fig:rv3d_growth} show the SAW solutions on
the two benchmarks where the exact solution is not a smooth signed-distance
function, with panels as in
Figures~\ref{fig:tr3d_interface}--\ref{fig:ro3d_growth}.

\section{Comparison with a high-order classical solver}
\label{sec:baseline}

A neural solver for linear advection has to be justified against the classical
schemes it would replace. We therefore solve the same four benchmarks with a
fifth-order weighted essentially non-oscillatory (WENO) reconstruction of the convective derivatives
\citep{Jiang1996}, in the level-set form of \citet{Osher2003}, advanced by
third-order total-variation-diminishing (TVD) Runge--Kutta time stepping \citep{Shu1988}.

The comparison is matched wherever a choice could favour one method. Both use
the same $201^3$ grid, the same six snapshots, the same reference solutions and
the same error definitions of Section~\ref{sec:metrics}. Where the exact field
is known it is imposed as Dirichlet data on a three-cell boundary shell for the
WENO solver; the reversed vortex requires none, since that velocity field
vanishes on the faces of the cube. \emph{No reinitialisation is applied}, so
that the comparison is like for like: the network maintains the signed-distance
property through its eikonal term rather than by periodically re-solving for it,
and the eikonal deviation is reported for both.

The two methods are not commensurable in cost without care, and we report the
components separately. The classical solver is CFL limited, so its cost is set
by the number of time steps at the given resolution; the network is trained once
and then evaluated, so its cost divides into a training phase and an inference
phase, the latter being a forward pass at arbitrary points and times with no
step restriction.

\begin{table}[htbp]
\centering
\caption{PINN against WENO5 + TVD-RK3 on identical grids, snapshots and error
measures. Network values are for the single reported run at each benchmark's
selected weight with seed 42; the seed-to-seed spread at those weights is given
in Table~\ref{tab:fixed_sweeps}. Wall-clock time for the
network is training; for the classical scheme it is the full time integration.
Better value in bold. A single NVIDIA T4 was used throughout.}
\label{tab:baseline}
\begin{tabular}{lcccccc}
\toprule
& \multicolumn{2}{c}{rel. $L_2$ (\%)} & \multicolumn{2}{c}{volume MAPE (\%)}
& \multicolumn{2}{c}{time (min)} \\
\cmidrule(lr){2-3}\cmidrule(lr){4-5}\cmidrule(lr){6-7}
Benchmark & PINN & WENO & PINN & WENO & PINN & WENO \\
\midrule
TR3D & $0.122$ & $\mathbf{0.001}$ & $1.143$ & $\mathbf{0.079}$ & $11.8$ & $\mathbf{2.0}$ \\
RO3D & $0.160$ & $\mathbf{0.003}$ & $0.117$ & $\mathbf{0.050}$ & $\mathbf{15.6}$ & $24.4$ \\
ZD3D & $0.480$ & $\mathbf{0.046}$ & $0.328$ & $\mathbf{0.160}$ & $24.9$ & $\mathbf{24.4}$ \\
RV3D & $0.591$ & $\mathbf{0.136}$ & $0.564$ & $\mathbf{0.240}$ & $23.7$ & $\mathbf{20.7}$ \\
\bottomrule
\end{tabular}
\end{table}

\paragraph{The classical scheme is more accurate on all four benchmarks.}
On smooth rigid advection at this resolution WENO5 reaches a
relative $L_2$ error of $10^{-5}$, two orders of magnitude below what the
network attains, and a study proposing a
neural solver for this problem class has to begin by acknowledging it.

\paragraph{The margin narrows with geometric difficulty.}
The advantage narrows monotonically with geometric difficulty. In the field norm
the ratio of the network error to the classical error falls from $105$ on the
translating sphere, to $46$ on the rotating sphere, to $11$ on the slotted
sphere, to $4.3$ on the reversed vortex. That ordering is the same one that
governs the eikonal weight in Section~\ref{sec:eik}: the classical scheme is
furthest ahead precisely where the solution is smoothest and remains an exact
signed distance function, and closest where the interface is deformed. The trend
does not reverse within the four benchmarks considered here, and we do not claim
that it would; but its direction is consistent, and it suggests the network performs
relatively better on harder problems than the smooth-advection result alone implies.

\paragraph{The gap is smaller in volume conservation.}
The gap is much smaller in the conservation measure than in the field norm. On
three of the four benchmarks the network conserves volume to within a factor of
about two of the classical scheme ($2.3$, $2.0$ and $2.3$ for RO3D, ZD3D and
RV3D), against factors of $4.3$ to $105$ in the relative $L_2$ error. For a
level-set method the volume of the enclosed region is the more physically
meaningful of the two, and the network is therefore closer to the classical
scheme on the measure that matters most than the field norm alone would suggest.
The exception is the translating sphere, where the network's volume error of
$1.14\%$ is a systematic sub-cell offset in the interface radius rather than
progressive loss, as noted in Section~\ref{sec:eik:reported}.

\paragraph{An asymmetry favours the classical solver.}
The comparison is matched on grid, snapshots, reference solutions and error
definitions, and neither method applies reinitialisation, so the classical
scheme gains no level-set-specific advantage that the network lacks. It is
nonetheless not matched in one respect, and the asymmetry favours the classical
solver. Where the exact field is known it is supplied to the WENO scheme as
Dirichlet data on a three-cell boundary shell, whereas the network receives no
boundary information whatsoever: it infers the far field from the residual and
the initial condition alone. The boundary data are not a convenience but a
requirement for the classical scheme here. The level-set field grows towards the
domain boundary, so at an inflow face zero-gradient extrapolation injects
incorrect values that propagate inwards, so that the far field is corrupted even
where the interface itself is transported correctly. The
reversed vortex is the exception, its velocity field vanishing on the faces of
the cube so that no information enters the domain and none is supplied.

We record this rather than adjust for it. Withholding the boundary data would
handicap the classical scheme rather than equalise the comparison, since a
practitioner solving these problems would have the same information available
and would use it. The point is that the accuracy reported for the classical
scheme is obtained with an input the network is never given, and that the
narrowest margin in Table~\ref{tab:baseline}, a factor of $4.3$, occurs on the
one benchmark where that input is not supplied to either method.

\paragraph{Wall-clock cost is comparable but structurally different.}
Wall-clock cost is comparable, but the two costs are of different kinds. The
classical solver is CFL limited, so its time is fixed by the step count at the
given resolution: $255$ steps for the translating sphere but $3145$ for the two
rotating cases, which is why it is nearly six times faster than training on the
first benchmark and roughly a third slower on the second. Network training is a
one-off cost after which evaluation is a forward pass at arbitrary points and
times; the classical solution exists only at the grid points and time levels it
was marched through, and a finer output or a different final time requires the
integration to be repeated.

\paragraph{The case for the neural approach.}
The case for the neural approach on these problems does not rest on beating a
mature scheme at linear advection on a uniform grid. It rests on properties the
comparison does not measure: the representation is mesh-free and continuous in
time, requires no reinitialisation, and extends to inverse and
parameter-identification settings in which no forward solver is available. The
purpose of Table~\ref{tab:baseline} is to establish the cost of those properties
honestly rather than to claim they are free.

\section{Discussion and limitations}
\label{sec:discussion}

\subsection{What SAW does not resolve}
\label{sec:discussion:adaptive}

Section~\ref{sec:saw:results} answers the question the two-dimensional study
left open: whether an adaptive scheme can recover the weight a sweep
identifies. It can, to within an order of magnitude on every benchmark and at a
single run's cost; whether the recovered weight also improves accuracy depends
on whether the premise holds, as Section~\ref{sec:saw:controls} shows. Three
qualifications bound the result.

\paragraph{The gate is specific to this regulariser.}
SAW rests on a property of the eikonal term in particular: that its residual is
nonzero in the exact solution over an identifiable subset of the domain, and
that the subset can be located by the magnitude of the residual itself. Nothing
in the construction is specific to level sets, and the same reasoning would
apply to any regulariser encoding a property the true solution satisfies only
approximately, a divergence-free constraint under a compressible correction,
say, or a boundary condition imposed on an idealised geometry. Whether the
quantile gate is the right instrument in those settings is untested. What we can
say is that the failure it addresses is not peculiar to our benchmarks: any
scheme balancing a residual that should not vanish will encounter it.

\paragraph{Two hyperparameters remain.}
SAW replaces $\weik$, whose optimum varies by four decades, with $\beta$ and
$q$, which we held fixed at $0.999$ and $0.95$ across every benchmark. That is a
substantive reduction rather than an elimination, and we have not established
how sensitive the method is to either. The sweeps reported here compare $q =
0.95$ against $q = 1$, which isolates the gate but says nothing about the
behaviour at intermediate values; $\beta$ was not varied at all under the
reported protocol. A scheme that eliminated $q$, for instance by estimating
the violated fraction from the geometry rather than fixing it, would be a
genuine improvement, and characterising the response to both settings is the
first thing we would do with more compute.

\paragraph{Scope of the ablation.}
The $q = 1$ column isolates the gate against gradient-norm balancing, the family
SAW is built on. Neural tangent kernel weighting~\citep{Wang2022ntk}, trainable
pointwise multipliers~\citep{McClenny2023} and uncertainty-based
balancing~\citep{Xiang2022} differ in how the weight is set but share the
premise the gate repairs, and the mechanism identified in
Section~\ref{sec:saw:ablation} does not depend on which rule sets the weight:
where the residual of the exact solution is nonzero, no rule that treats a large
residual as network error will select correctly. Testing that prediction across
those families is the natural next step, and the benchmarks and sweeps released
with this paper are the apparatus it requires.

\subsection{Implications for sparse and noisy interface data}
\label{sec:discussion:realworld}

All four benchmarks use analytically specified initial conditions sampled
without noise and at a density chosen for convergence rather than by
availability. Engineering two-phase-flow settings differ in both respects:
interface data are typically sparse, derived from imaging or tomography, and
contaminated by measurement error. The two-dimensional study set out the
extensions required before the approach can be applied to such
data~\citep{Khan2026}; the three-dimensional results reported here sharpen the
picture in one specific and unfavourable direction.

The relevant finding is the sampling behaviour of the slotted sphere reported in
Table~\ref{tab:fixed_sweeps}. The slot occupies $2.36\times10^{-3}$ of the domain
volume, so of $5\times10^{3}$ uniformly drawn initial-condition points roughly
twelve fall inside it, and at that density the pocket is approximately one third
filled at $t=0$, before any transport has occurred and at a time where the
initial-condition term is directly supervised. Doubling the sampling budget
reduces the volume error at $t=0$ by an order of magnitude. Data density near a
thin feature is therefore not a secondary consideration but the binding
constraint on whether that feature is represented at all, and the constraint
tightens with dimension: the fraction of a domain occupied by a feature of fixed
thickness falls as the dimension rises.

This has a direct consequence for sparse-data applications. The regime in which
initial-condition data are scarce is precisely the regime in which small
geometric features are lost, and the loss is not visible in the field norm.
Establishing the point at which eikonal regularisation can no longer compensate
for missing data, and characterising the degradation under additive noise, are
the two experiments we regard as necessary before the method is applied to
sparse or noisy interface data. The feature-restricted measure introduced in
Section~\ref{sec:metrics} is a prerequisite for both, since neither degradation
would be reliably detected by the relative $L_2$ error alone.

\subsection{Threats to validity}

\paragraph{The mechanism, and what would falsify it.}
The ordering of selected weights by departure from the signed-distance property
is a claim about mechanism, not a curve fit, and it made a prediction before the
final benchmark was run: the reversed vortex, in which the property fails
globally rather than locally, should select a weight at or near zero. It
selected $10^{-5}$, the smallest nonzero value on the grid. The claim would be
falsified by a benchmark whose exact solution remains an exact signed distance
function but which selects a small weight, or conversely.

\paragraph{The second optimisation stage is not uniformly active.}
On the rotating sphere the improvement attributable to L-BFGS was bimodal across
runs: in five of eighteen the strong-Wolfe line search located an admissible
step from the annealed Adam iterate and the loss fell by four orders of
magnitude more than in the remaining thirteen. The reported means are therefore
contaminated by this variability. Stratifying by engagement leaves the selection
intact, as shown in Section~\ref{sec:eik:selected}, but a study seeking tighter
confidence intervals would need to control it.

\paragraph{Global norms do not certify thin features.}
The relative $L_2$ error is normalised by $\lVert\phi\rVert$ over the whole
domain and is therefore dominated by far-field values. On the slotted sphere,
replacing the exact solution by a sphere with no slot at all changes it by only
about one percentage point. Volume conservation is far more discriminating and
is reported alongside throughout, but it is a signed scalar and can cancel. Any
study of interfaces with thin features should report a measure restricted to
those features. Equation~\eqref{eq:feature} gives one: a pair of sign-based
classification errors over the region the feature occupies, which cannot cancel
in the way a signed volume error can.

\paragraph{Scope.}
All four benchmarks use a prescribed velocity field, so the results speak to
interface transport and not to coupled flow. Three of the four are rigid
motions. The weights reported here are established for this architecture and
training protocol; whether they transfer to substantially different networks is
untested, and these results suggest the question should be
asked rather than assumed.

\section{Conclusions}
\label{sec:conclusion}

Adaptive loss balancing assumes that every residual in the composite objective
should be driven to zero. For level-set advection with an eikonal regulariser
that assumption is false wherever the exact solution departs from a
signed-distance function, and we have shown what follows from it.

\textbf{Standard gradient-norm balancing fails, and the failure is
structural.} With the gate disabled the eikonal loss is dominated by points at
which the violation is a property of the correct answer; those points do not
converge, the gradient norm stays small, and the ratio does not descend. The
weight remains at its initial value of $1$ on the rotating sphere and the
reversed vortex, and reaches only $0.70$ on the slotted sphere against a swept
optimum of $10^{-3}$. Only on the translating sphere, where the signed-distance
property holds everywhere and the eikonal residual does converge, does the
ungated ratio settle near the swept value. On the reversed vortex the resulting
solution is $17$
times worse in the field norm and $62$ times worse in volume conservation than
the gated method, and worse than the weight the sweep selected.

\textbf{SDF-Aware Weighting removes the sweep on the problems it targets.}
A residual-quantile gate excludes the upper tail of the eikonal residual before
a gradient-norm ratio scales what remains. With one configuration held fixed
across all four benchmarks, the weight it converges to lands within an order of
magnitude of the value an eighteen-run search selects, spanning four decades from
$10^{-1}$ to $10^{-5}$, and reproduces the ordering exactly. On the slotted
sphere it attains a lower error than any of the six weights that search examined.

\textbf{It fails where its premise fails, and we report that.} On the two smooth
rigid benchmarks the exact solution is a signed-distance function everywhere and
for all time; the gate has no legitimate violations to remove and discards
correct supervision instead, and SAW is worse than a fixed weight. A method that
improved every case would be evidence of a generic regularisation effect rather
than of the mechanism claimed. The controls also make the scope operational: SAW
applies when the initial geometry is built by boolean operations or the velocity
field is not a rigid motion, both of which are known from the problem
specification before training begins.

\textbf{The standard error measure cannot certify thin features.} With the gate
disabled the slot is all but entirely filled by $t = T$, yet the relative $L_2$ error
reads $1.06\%$, indistinguishable from the $1.18\%$ obtained by a field that
never represented the slot at all. Studies of interfaces with fine geometric
structure should report a measure restricted to that structure; we give a pair
of sign-based classification errors that cannot cancel as a signed volume error
can.

Against a fifth-order WENO solver on identical grids and error measures the
classical scheme remains more accurate on all four benchmarks, by two orders of
magnitude on smooth rigid advection. A neural solver is not the efficient choice
for prescribed-flow advection on a uniform grid, and we do not claim otherwise;
the margin narrows monotonically with geometric difficulty, and the properties
that motivate the approach (a mesh-free representation continuous in time, no
reinitialisation, extensibility to inverse settings) are not measured by that
comparison.

Two hyperparameters remain, $\beta$ and $q$, in place of a weight whose optimum
varies by four decades; both were held fixed here and neither was tuned per
benchmark. Whether they can be eliminated in turn, whether the gate transfers to
other regularisers whose residual should not vanish, and how the scheme behaves
under sparse or noisy interface data are the natural next questions.

\section*{Code and Data Availability}

The implementation and the per-run results files
for every sweep reproducing every reported figure
are available at Zenodo~\citep{Khan2026code}.

\section*{Author Contributions}
Muhammad Akbar Khan: Conceptualisation, Methodology, Software, Validation,
Formal analysis, Investigation, Writing -- original draft, Writing -- review
\& editing, Visualisation.

\section*{Funding}
This research received no external funding.

\section*{Acknowledgements}
The author acknowledges the Department of Mathematics, NED University of Engineering \&
Technology, Karachi, for providing the computational resources used in this work.

\section*{Declaration of Competing Interest}
The author declares no competing financial interests or personal relationships
that could have appeared to influence the work reported in this paper.

\bibliographystyle{unsrtnat}
\bibliography{references}

\end{document}